\documentclass[preprintnumbers,reprint,twocolumn,superscriptaddress, amsmath,amssymb,aps,nofootinbib]{revtex4-1}
\usepackage{graphicx}
\usepackage{dcolumn}
\usepackage{bm}
\usepackage{float}
\usepackage{hyperref}
\usepackage{dsfont}
\usepackage{slashed}
\usepackage{color}
\usepackage{amsmath}
\usepackage[section]{placeins}
\usepackage{braket}
\usepackage{upgreek}
\usepackage[bottom]{footmisc}
\usepackage[normalem]{ulem}

\newcommand{\beq}{\begin{eqnarray}}
\newcommand{\eeq}{\end{eqnarray}}
\newcommand{\beqnn}{\begin{eqnarray*}}
\newcommand{\eeqnn}{\end{eqnarray*}}

\newcommand{\tr}{\text{Tr}}

\def\bs{\boldsymbol} 
\def\del{\partial}

\newcommand{\eqn}[1]{Eq.~\eqref{#1}}

\long\def\comment#1{ }

\def\be{\begin{eqnarray*}}
\def\ee{\end{eqnarray*}}
\def\beq{\begin{eqnarray}}
\def\eeq{\end{eqnarray}}

\def\rme{{\rm e}}
\def\rmd{{\rm d}}

\def\coll{{\rm c}}
\def\cs{{\rm cs}}
\def\Gl{{\rm G}}

\def\hc{{\rm hc}}
\def\out{{\rm loss}}
\def\med{{\rm med}}

\def\cS{{\cal S}}
\def\cH{{\cal H}}

\def\bfk{{\bs k}}
\def\bfq{{\bs q}}
\def\bfu{{\bs u}}
\def\bfr{{\bs r}}
\def\bfo{{\bs 0}}
\def\bfx{{\bs x}}

\begin{document}

\title{Factorization for jet production in heavy-ion collisions}

\author{Yacine Mehtar-Tani}
\email{mehtartani@bnl.gov}
\affiliation{Physics Department, Brookhaven National Laboratory, Upton, NY 11973, USA}

\author{Felix Ringer}
\email{felix.ringer@stonybrook.edu}
\affiliation{Center for Nuclear Theory, Department of Physics and Astronomy,
Stony Brook University, NY 11794, USA}
\affiliation{Thomas Jefferson National Accelerator Facility, Newport News, VA 23606, USA}
\affiliation{Department of Physics, Old Dominion University, Norfolk, VA 23529, USA}

\author{Balbeer Singh}
\email{Balbeer.Singh@usd.edu}
\affiliation{Department of Physics, University of South Dakota, Vermillion, SD 57069, USA}

\author{Varun Vaidya}
\email{Varun.Vaidya@usd.edu}
\affiliation{Department of Physics, University of South Dakota, Vermillion, SD 57069, USA}

\preprint{JLAB-THY-24-4137}


\begin{abstract}
We develop an Effective Field Theory approach for jet observables in heavy-ion collisions, where the jet is treated as an open quantum system interacting with a hot and dense QCD medium. Within this framework, we derive a novel factorization formula for inclusive jet production, expressed as a series expansion with an increasing number of radiating subjet functions that encode forward scattering with the Quark-Gluon Plasma, convolved with perturbative matching coefficients. This work provides a systematic framework for computing jet observables at higher order and understanding their non-perturbative aspects, paving the way for future applications in heavy-ion phenomenology. 
\end{abstract}

\maketitle
In his seminal work \cite{Bjorken:1982tu}, J. D. Bjorken hypothesized that a transient quark-gluon plasma (QGP) could be produced in proton-proton ($pp$) collisions alongside high-energy quark and gluon jets, which may lose energy as they traverse the QGP. This remarkable phenomenon of ``jet quenching'' was observed in Au-Au collisions at RHIC and later at the LHC, spurring significant experimental and theoretical efforts to investigate the properties of the newly discovered state of matter~\cite{BRAHMS:2004adc,PHOBOS:2004zne,STAR:2005gfr,PHENIX:2004vcz,ATLAS:2010isq,
ALICE:2010yje,
CMS:2011iwn,ATLAS:2018gwx,CMS:2021vui,ALICE:2023waz,Connors:2017ptx}.

However, the theory of jet quenching has yet to attain the predictive power of simpler systems, such as electron-positron or $pp$ collisions and remains a fundamental challenge for many-body QCD. The purpose of this letter is to advance this goal by constructing a systematic framework for jet observables in ultra-relativistic heavy-ion collisions (HIC) using the key idea of factorization. The study of jets in $pp$ collisions has achieved unprecedented quantitative precision with QCD calculations~\cite{Currie:2016bfm,Czakon:2019tmo} and Effective Field Theory (EFT) tools~\cite{Larkoski:2017jix,Asquith:2018igt,Marzani:2019hun} such as Soft Collinear Effective Theory (SCET) \cite{Bauer:2002aj,Bauer:2003mga,Bauer:2000yr,Bauer:2001ct,Bauer:2002nz}. This has been achieved by using factorization formulas~\cite{Collins:1989gx,Larkoski:2017jix}, that separate perturbative dynamics at short distances from non-perturbative physics at long-distance scales encoded in universal functions such as parton distribution functions (PDFs) for the initial state or shape functions~\cite{Korchemsky:1999kt,Lee:2006nr} 
related to hadronization that can be extracted from reference processes and lattice data. 

An important phenomenon in partonic energy loss is the Landau-Pomeranchuk-Migdal (LPM) effect \cite{Landau:1953um,Migdal:1956tc}, which is a coherent action of multiple scattering centers that causes energetic color charges to lose energy to the plasma. This was first understood in the 1990s~\cite{Gyulassy:1993hr,Wang:1994fx,Baier:1994bd,Baier:1996kr,Baier:1996sk,Zakharov:1996fv,Zakharov:1997uu,Gyulassy:2000er,Wiedemann:2000za,Guo:2000nz,Wang:2001ifa,Arnold:2002ja,Arnold:2002zm} and phenomenological models were developed to describe the data~\cite{Salgado:2003gb,Liu:2006ug,Qin:2007rn,Armesto:2011ht}. However, it was realized that, because jets are extended multi-partonic systems, interference between multiple fast-moving color charges occurs in the plasma, which depends on the resolution power of the QGP. Any complete theory needs to account for this interference-driven phenomenon, known as color (de)coherence~\cite{Mehtar-Tani:2010ebp,Mehtar-Tani:2012mfa,Casalderrey-Solana:2011ule, Casalderrey-Solana:2012evi} as well as the LPM effect. 

In this letter, we focus on inclusive jet production, which is crucial for future extensions of the EFT to jet substructure observables~\cite{Singh:2024vwb}. The relevant cross section is a histogram of jets based on their transverse momentum $p_T$ and rapidity $\eta$, for a fixed jet radius $R$.  We derive a factorization formula for this observable that, for the first time, allows a clear separation between vacuum and medium physics at all orders in perturbation theory while accounting for interference phenomena. We further show how the in-medium jet evolution can be factorized from the universal observable-independent dynamics of the medium. This is a significant extension of previous results in the literature where EFT techniques were applied to jets in HIC ~\cite{Idilbi:2008vm,DEramo:2010wup,Ovanesyan:2011xy}.  We work in the narrow-jet $R\ll 1$ limit, which has been studied in $pp$ collisions and is a good approximation even for relatively large values of $R$~\cite{Dasgupta:2014yra,Kaufmann:2015hma,Kang:2016mcy,Dai:2016hzf,vanBeekveld:2024jnx}. Extending the vacuum factorization to include medium interactions requires us to treat the jet as an open quantum system, with several new explicit and emergent scales that we discuss below.

{\it The EFT landscape.} In vacuum, the inclusive jet cross section involves two scales -- the transverse momentum $p_T$ ($100-1000$~GeV) and the jet scale $p_T R$. For $R\ll 1$, the separation of these scales allows a factorization of the cross section in terms of hard and jet functions. The resummation of logarithms in the expansion parameter $R$ is achieved via a DGLAP-type evolution equation \cite{Dasgupta:2014yra,Kang:2016mcy,Dai:2016hzf}. The presence of the QGP introduces new scales such as the medium length $L$, temperature $T$, Debye mass $m_D\sim gT$, where $g$ is the QCD coupling, and the mean free path of the jet partons $\ell_{\rm mfp}$. However, the medium dependence of jet observables is mainly encoded in a single emergent scale, which is associated with the transverse momentum gained by a parton, $Q^2_{\rm med} \equiv \langle k_\perp^2\rangle \sim \hat q$, where $\hat q$ is the jet transport parameter and $L$ the medium length \cite{Baier:1996sk}.  
Another scale is the coherence angle $\theta_c \sim 1/(Q_{\rm med}L)$, which controls color decoherence in the medium. This is the minimum angular separation for resolving the color of partons~\cite{Casalderrey-Solana:2014bpa}.  

At current collider experiments, the medium temperature is in the range of $0.5-1$~GeV, which is a non-perturbative scale. As a result $g\sim \mathcal{O}(1)$ and, $m_D \sim T$. We denote the jet energy loss scale, i.e., the energy transported out of the jet by $E_\out$. If we model this effect on the differential cross section as a shift of the jet $p_T$ by $E_\out$, the nucleus-nucleus $AA$  spectrum behaves approximately as  
$\rmd \sigma_{ AA}/ \rmd p_T \propto (p_T+E_\out)^{-n}$. The nuclear modification factor, which is the ratio of the $AA$ to $pp$ cross section can then be approximated by, $R_{AA}\sim (1+E_\out/p_T)^{-n}\simeq 1 - n \,E_\out/p_T+\ldots\;$. Since, $R_{AA} \lesssim 1$ and $n \sim 5$, we have  $E_\out \lesssim p_T/n \ll p_T$, leading to the following hierarchy of scales
\beq 
p_T  \,\gg \, p_TR \, \gtrsim\, E_\out\, \gg  \,Q_{\rm med}  \,\gtrsim  \,T  \, \sim   \, \Lambda_{\rm QCD}\,.
\eeq
There is another scenario where $p_T \gg p_TR \sim Q_{\text{med}}$, where the virtuality of the jet $\sim \mathcal{O}(Q_{\rm med})$. Observables with this hierarchy were discussed in the Refs.~\cite{Vaidya:2020cyi,Vaidya:2020lih,Vaidya:2021vxu,Vaidya:2021mly}, which is relevant for low-energy or ultra-narrow jets.

\begin{figure}[t]
\centering
\includegraphics[scale=0.19]{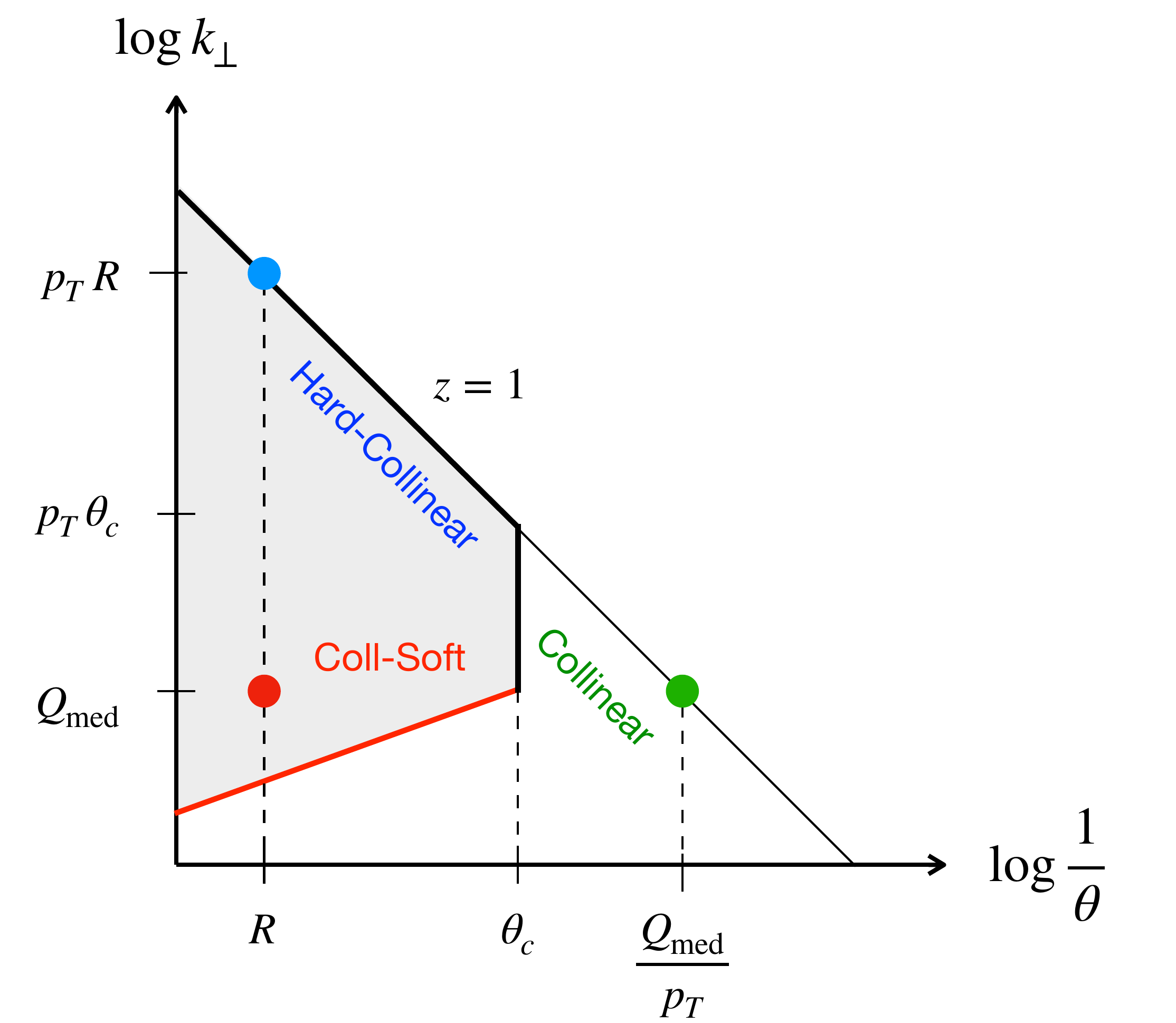}
\caption{Lund plane with the measurement and kinematic constraints for inclusive jet production in HIC. The colored points at the intersections of different lines represent modes in the EFT~\label{fig:lund} (See main text for details).}
\end{figure}

Along with the jet radius $R$, we can define two expansion parameters $\beta \equiv  E_\out/p_T$ , and $\delta \equiv Q_\med /E_\out$. Any loss of energy through radiation can only occur at angles larger than $R$, implying $E_{\out} \lesssim Q_{\rm med} /R$ so that $ \delta \sim R$. Furthermore, we will focus on the hierarchy $\beta \sim R$, which is relevant to phenomenological studies. We choose a frame where the jet is propagating along the $z$-axis, i.e. in the light-like direction $n^{\mu}\equiv (1,0,0,1)$. In what follows, we adopt the light-cone decomposition of four-momenta $p^{\mu}=  p^-\, \, n^{\mu}/2+p^+\,\bar n^{\mu}/2+p_\perp^{\mu }$, where $\bar n^\mu \equiv(1,0,0,-1)$. To illustrate the phase space of radiation contributing to our observable, we use the Lund plane~\cite{Andersson:1988gp} representation, see Fig.~\ref{fig:lund}. It displays the transverse momentum $k_\perp=z\theta p_T$ of emissions as a function of their angle $\theta$, where $z$ is the momentum fraction of the initial parton that ends up in the observed jet. From the intersection of the energy constraint $z=1$ and the radius $R$, we can identify a hard-collinear (hc) mode (blue dot) with momentum scaling $p^{\mu}_\hc\equiv (p_\hc^-,p_\hc^+, p_\hc^{\perp}) \sim p_T(1,R^2,R)$.
The medium imparts a small transverse momentum  $Q_{\text{med}}$ to the jet partons. As shown in Fig.~\ref{fig:lund}, the $k_{\perp} \sim Q_{\rm med}$ line intersects the phase space constraints, $\theta\sim R$ and $z \sim 1$, at two distinct points. Hence, we identify two additional modes: a collinear-soft (cs) mode with momentum scaling 
$p_\cs^{\mu}\sim p_T\beta(1, \delta^2 ,\delta)$ where $\delta \sim R$ and a collinear mode $p_\coll^{\mu}\sim p_T(1,\delta'^2,\delta' )$, where $\delta'\equiv \beta \delta = Q_{\rm med}/p_T \sim R^2$. The formation time for a radiated parton is given as $t_F=zp_T/k_{\perp}^2$ and defines a time scale over which interactions with the medium remain coherent. The condition $t_F \sim L$ sets an additional boundary shown by the red line in Fig.~\ref{fig:lund}. Modes to the right of this line interact coherently with the medium leading to an LPM suppression. Hence, the collinear medium-induced radiation will be be suppressed. Another boundary, identified in \cite{Mehtar-Tani:2017web,Caucal:2018dla}, corresponds to the relation $k^2_\perp\sim \hat q t_F$, which will be addressed in future work. Finally, we show the coherence angle $\theta_c$. Each resolved hc parton within the jet acts as a source of cs radiation. The collinear radiation remains unresolved by the medium and does not contribute to the cross section and will be omitted in the subsequent discussion.

The shaded gray region in Fig.~\ref{fig:lund} is the relevant phase space of emissions that contributes to our observable at leading power in $R$, $\delta \sim R$, and $\beta \sim R$, which is captured by two modes -- the hc and the cs mode. The medium-induced cs radiation has much lower energy than the hc mode and its contribution to jet energy loss starts at ${\cal O}(\beta)$. As a result, relative to the vacuum evolution, medium effects are power suppressed but may become a leading power contribution when $\beta\sim 1/n$, as discussed above. The medium is made up of soft partons, which uniformly populate all momentum directions in the rest frame with momentum scaling $p_{\rm s}^{\mu}=p_T(R^2,R^2,R^2)$. They act as sources for Glauber gluons that scale as $p_\Gl^{\mu}=p_T(R^2,R^3,R^2)$ and mediate forward $t$-channel scattering with the cs partons.

{\it Factorization.}
We now describe a multi-stage factorization formula for the inclusive jet cross section in HIC. A complete technical derivation of the results presented hereafter will be given in a longer companion paper \cite{LongDerive}.
\begin{figure}[t]
 \centering
\includegraphics[scale=0.175]{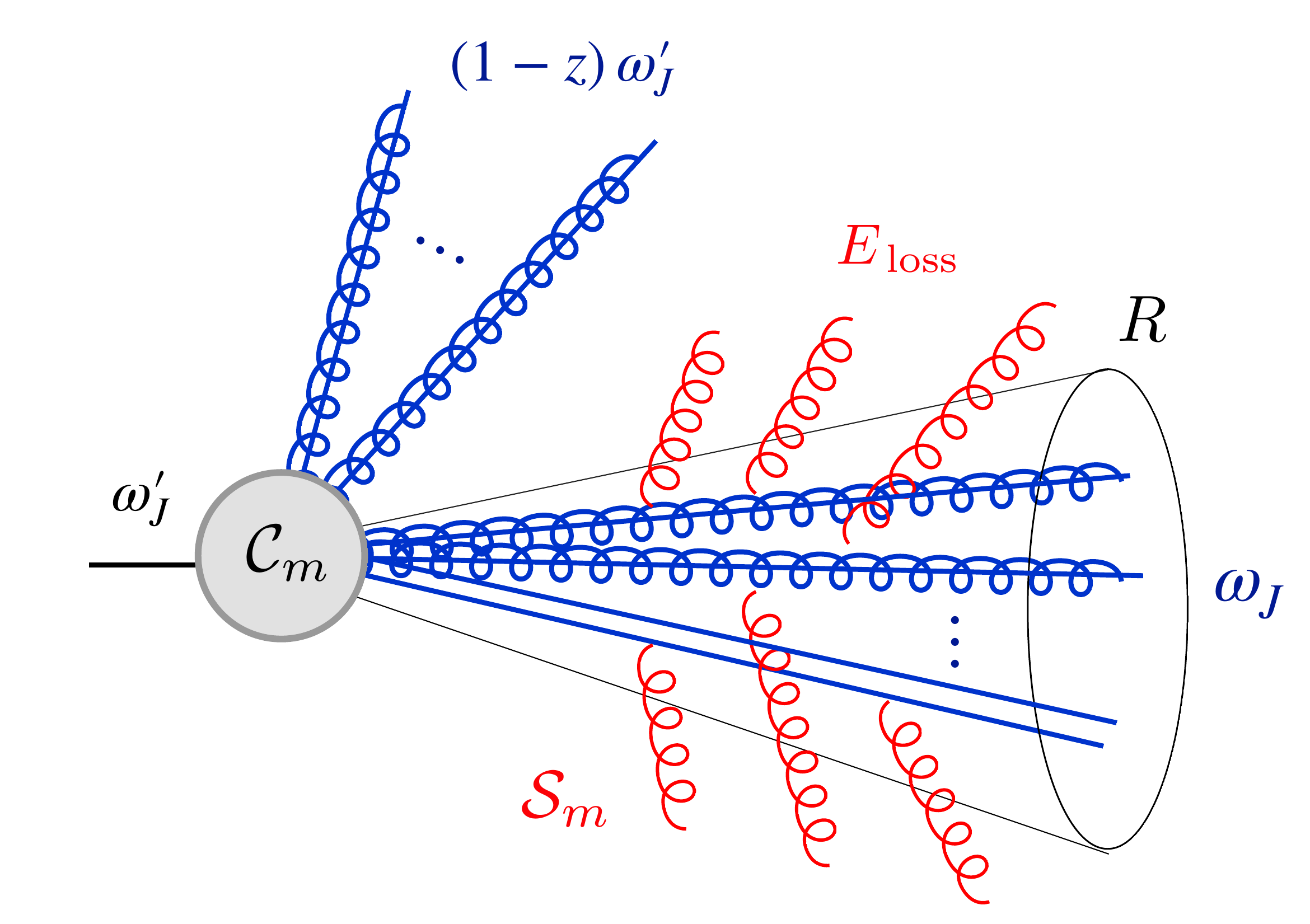}
 \caption{Illustration of the jet cone and the various color-coded modes that contribute to the factorized inclusive jet cross section in heavy-ion collisions.~\label{fig:thetac}}
\end{figure}
 
{$\bullet$ Hard to hard-collinear regime:  }
We first integrate out modes with virtuality larger than $Q\sim p_T R$. This allows us to factorize the hard-scattering matrix elements, where the jet is initiated at time scales $p_T^{-1}$, from its subsequent collinear evolution in the vacuum and medium at times larger than $p_T/Q^2\sim (p_T R^2)^{-1}$. At this stage, we identify a single hc mode $p_{\rm hc}^\mu$ accounting for fluctuations down to the medium scale $T$. This mode also includes the cs, soft, and  Glauber contributions described above. Thus, the inclusive jet cross section becomes
\begin{equation}
\!\!\! \frac{{\rm d}\sigma}{{\rm d}p_T{\rm d}\eta}= \sum_{i \in q, \bar q , g}\int_0^1 \frac{{\rm d}z}{z} \, H_i\left(\omega= \frac{\omega_J}{z},\mu\right) \, J_i(z, \omega_J,\mu)\,.
\label{eq:factI}
\end{equation}
Here, $H_i$ are the hard-scattering functions including also PDFs, which describe the production of a massless parton $i$ with four-momentum $p^\mu$ and  $ \omega\equiv p^- $ denotes its large comentum component.  The functions $J_i$ describe the jet evolution where $z=\omega_J/\omega$ is the momentum fraction of the jet initiating parton $i$ that ends up in the measured jet. The momentum fraction $z$ is related to the jet $p_T$ and rapidity $\eta$ via $\omega_J = 2p_T \cosh \eta$.  The jet function for a quark is 
\begin{align} 
J_q(z, \omega_J,\mu)\equiv \frac{z}{2N_c} \int {\rm d} s\, {\rm e}^{is \omega} \text{Tr}\Big[\bar{\chi}_{n}(0)\rho_M \frac{\slashed{\bar n}}{2} \chi_{n} (s \bar n)\mathcal{M}\Big].
 \label{eq:Sigma}
\end{align}
nHere, $\chi_n$ is the $n$-collinear hc Dirac field dressed with a collinear Wilson line $W_n$ defined in Appendix \ref{app:Ocs}. Moreover, $\rho_M$ is the initial medium density matrix, $\mathcal{M}$ is the measurement function defined as $\mathcal{M} = \delta(z - \omega_J/\omega)\Theta_{\text{alg}}$, where $\Theta_{\rm alg}$ denotes phase space constraints due to the jet clustering algorithm. The trace is over color and Dirac indices as well as all final states. The gluon jet can be defined similarly, see Appendix \ref{app:Ocs}. The physics of the medium is contained in the jet function and it obeys the same DGLAP-type evolution equation as in vacuum~\cite{Dasgupta:2014yra,Kaufmann:2015hma,Kang:2016mcy,Dai:2016hzf,vanBeekveld:2024jnx}. See also Refs.~\cite{Kang:2017frl,He:2018gks,Qiu:2019sfj}.

{$\bullet$ Hard-collinear to collinear-soft regime:  }
Next, we match the hard-hc factorization to an EFT that only contains modes with virtuality at the medium scale $Q_{\text{med}}$. In the matching, we only keep terms at leading power in $Q_{\text{med}}/(p_TR)$, i.e., we drop hard interactions with momentum exchange $k_{\perp}\equiv|\bfk|\sim p_TR \gg Q_{\text{med}}$, which are power suppressed. The physical picture is that the vacuum evolution of the jet leads to a high-energy collinear parton cascade, producing a collection of hard subjets or prongs that propagate through the medium. Each subjet (hc mode) defines a new direction inside the jet and acts as a source for cs radiation as shown in Fig.~\ref{fig:thetac}. The physics of color decoherence is encoded in the interference of the cs radiation sourced by these subjets. We split the hc field into three modes: the hard-collinear,  collinear-soft, and soft modes $ A^\mu_{\rm hc} \to A_{\rm hc}^\mu + A_{\rm cs}^\mu+A_{\rm s} $ 
and  $ \chi_{\rm hc} \to \chi_{\rm hc} + \chi_{\rm cs}+\chi_{\rm s}$ 
with the respective momentum scalings as introduced in the previous section. Therefore, we refactorize the jet function in Eq.~(\ref{eq:Sigma}) in terms of a series with an increasing number of cs subjet functions convolved with perturbative matching coefficients
\begin{widetext}
\begin{align}
\label{eq:J-S-fact}
&  J_i(z,\omega_J,\mu) =\int_{0}^{1} \rmd z' \,\int_0^{+\infty}\rmd \epsilon \, \delta(\omega_J'-\omega_J-\epsilon)\, \sum_{m}\prod_{j=2}^m\int \frac{{\rm d}\Omega(n_j)}{4\pi} \mathcal{C}_{i\rightarrow m}\Big(\{\underline{n}\},z', \omega_J'= \frac{z'\omega_J}{z}, \mu,\mu_\cs\Big)\, {\cal S}_{m} (\{\underline{n}\} , \epsilon,\mu_\cs)  \,, 
\end{align}
which holds up to power corrections ${\cal O}(Q_{\rm med}/(p_TR))$, Here, $\mu$ and $\mu_\cs$ are the hard and collinear-soft factorization scales, respectively. In SCET, one usually sets $\mu_\cs=\mu$.  The cs functions ${\cal S}_{m}$ are defined as 
\begin{align}
    \label{eq:soft-funct}
&{\cal S}_{m}(\{\underline{n}\},\epsilon,\mu) \equiv    \text{Tr}\Big[U_m(n_m)...U_{1}(n_1)U_0(\bar n)\rho_M U^\dag_0(\bar n)U_1^\dag(n_1)...U_m^\dag(n_m)\mathcal{M} \Big] \,,
\end{align}
\end{widetext}
where $\{\underline{n}\}\equiv\{n_1,n_2,...,n_m\} $ denotes the directions of the $m$ collinear subjets. Additionally, we have $n_i\cdot n_j\sim R$ for $i\neq j$ and $\mathcal{M}= \Theta_{\text{alg}}\delta(\epsilon - \bar n \cdot p_\out )$ where $\epsilon$ is the energy loss due to radiation escaping the jet. The Wilson line $U_0(\bar n)$, which ensures gauge invariance, describes an unresolved effective charge moving in the opposite direction. The path ordered Wilson line reads 
\beq 
U(n)\equiv {\bf P} \exp\left[ig \int_0^{+\infty} \rmd s  \, n\cdot A_{\rm cs}(s n)\right]\,.
\label{eq:CSWilson}
\eeq
The perturbative coefficient $\mathcal{C}_{i\rightarrow m} $ starts at ${\cal O}(\alpha_s^{m-1})$ and describes the vacuum production of $m$ partons inside the jet at pairwise angles $\theta\gg  Q_\med/p_T\sim R^2$ from the initiating parton $i$. This refactorization takes the same form as encountered in the context of non-global logarithms in Ref.~\cite{Becher:2015hka}. See also ~\cite{Dasgupta:2001sh,Larkoski:2015zka}.\\

{\it Unresolved jet production.} 
We start by focusing only on the first term of the series in Eq.~(\ref{eq:J-S-fact}), which corresponds to jets that are not resolved by the medium, i.e., $\theta_c \gtrsim R$. An analysis for $ \theta_c \ll R$ will be left for future work. A calculation to ${\cal O}(\alpha_s)$ confirms the consistency between the anomalous dimensions of the various functions in~\eqn{eq:J-S-fact}, which we discuss in more detail in Appendix \ref{app:RG}. In particular, the anomalous dimensions of the functions $\mathcal{C}_{i\rightarrow m} $ and ${\cal S}_m$ add up to the DGLAP splitting functions, see Eq.~(\ref{eq:factI}). 

Our ultimate goal is to separate the collinear-soft physics of the jet from the universal soft physics of the medium. The two corresponding modes have the same virtuality but are separated in terms of their rapidity. Therefore, we have split the gauge field into ``fast'' (collinear-soft) and ``slow'' (soft) modes: $A^\mu_{\rm cs}(p^->\nu)+A^\mu_{\rm s}(p^-<\nu)$, where $\nu$ is a rapidity factorization scale \cite{Balitsky:1995ub,Chiu:2012ir}. The field describing the fast modes $A^\mu_{\rm cs}$ can then be viewed as a quantum fluctuation propagating in the presence of the background of slow fields $A^{\mu}_s$~\cite{Iancu:2014kga,Blaizot:2012fh,Blaizot:2014bha,Blaizot:2013vha}. 

In this work, we treat the scattering centers, namely, the soft partons, to be independent or color uncorrelated. This is valid when $\ell_{\rm mfp}$ is significantly larger than the color coherence length $\xi_D \sim 1/m_D$ in the medium, i.e., $\ell_{\rm mfp} \gg \xi_D$.
To obtain a factorization, we write the single subjet function as a series in the number of jet-medium interactions ${\cal S}_{1} = \sum_{n=0}^{\infty}{\cal S}_{1}^{(n)}$, see Eq.~(\ref{eq:soft-funct}). Here, ${\cal S}_{1}^{(0)}$ is the vacuum contribution. At ${\cal O}(n)$, 
we have
\begin{align}\label{eq:FctAl}
&{\cal S}_{1}^{(n)}(\epsilon,\mu_\cs)\! =\! (8\pi \alpha_s)^{2n}\Bigg[\prod_{i=1}^{n}\int_{0}^{\infty}\! \!{\rm d}x^-_i\Theta(x^-_i-x_{i+1}^-)\!\int\! \frac{{\rm d}^2\bfk_{i}}{(2\pi)^3}\nonumber \\
&\varphi(\bfk_{i},\mu_\cs,\nu,x^-_i)\Bigg]
{\bf F}_{1}^{(n)}(\epsilon; \bfk_{1}, \ldots, \bfk_{n}; x_1^-, \ldots x_n^-;\mu_\cs,\nu)\,.
\end{align}
\normalsize
The $n=0$ term corresponds to the vacuum contribution ${\cal S}_{1}^{(0)}={\cal S}^{\rm vac}_1$, with its one-loop expression provided in the Appendix \ref{app:BFM}. 
This expression contains a product of $n$ copies of the medium correlator $\varphi$, which only contains soft physics ($p^- <\nu$) and is defined as
\begin{align}
\label{eq:MedFn}
&\varphi(\bfk,\mu,\nu,x^-)= \frac{1}{\bfk^2}\frac{1}{N_c^2-1}\int \frac{{\rm d}k^-}{2\pi}\int {\rm d}^4r\, e^{-i \bfk \cdot \bfr+ik^-r^+}\nonumber\\
&\text{Tr}\Big[O^a_{\rm s}(\bfr,r^+,r^-+x^-) \rho_M O^a_{\rm s}(\bfo,0^+,x^-)\Big]\,,
\end{align}
where
$\mathcal{O}_{\rm s}^{a}=\bar{\chi}_{\rm s}S_n t^a\frac{\slashed{n}}{2}S_n^{\dagger}\chi_{\rm s}$, for a quark,  with the soft Wilson line $ S_{n} =  \ {\bf P} \exp \{ig\int_{-\infty}^0 {\rm d}s\, n\cdot A_{\rm s}(x+s n)\}$. 
The $x^-$ dependence of $\varphi$ is encoded in the density matrix that is assumed to be slowly varying $x^- \gg r^-$. This factorization can be achieved by using the SCET-based formulation developed in Ref.~\cite{Rothstein:2016bsq} where the Glauber modes are explicitly sourced by on-shell propagating soft fields of the medium. The off-shell Glauber mode is integrated out resulting in effective $t$-channel exchange operators between the soft and cs partons at the level of the action, at leading order in $\delta$. The cs dynamics are described by the standard collinear SCET Hamiltonian. The interaction Hamiltonian describing forward scattering of a cs gluon off a soft medium parton, mediated by a single Glauber exchange and medium-induced cs radiation to all orders in $\alpha_s$ can be recovered from Ref.~\cite{Rothstein:2016bsq}. See Appendix \ref{app:Ocs}. 

The correlator $\varphi$ depends only on the universal microscopic properties of the medium. It evolves with the full QCD Hamiltonian and a similar object appears in the context of saturation~\cite{Stewart:2023lwz}. Here, $\varphi$ obeys the Balitsky-Fadin-Kuraev-Lipatov (BFKL) equation \cite{Balitsky:1978ic,Kuraev:1977fs} in the rapidity scale, see Appendix \ref{app:RG}. The function ${\bf F}_{1,q}^{(n)}$ in Eq.~(\ref{eq:FctAl}) obeys a BFKL equation in all its arguments $\bfk_{i}$ but with opposite sign compared to $\varphi$. Likewise, it also has the threshold anomalous dimension $\gamma_{{\cal S}_1}(z)$ ( Eq.~(\ref{eq:gammaS} in Appendix \ref{app:RG}), which is associated with the scale $\mu_\cs$. This maintains renormalization group (RG) consistency in terms of the scales $\nu$ and $\mu_\cs$, respectively. The dense regime leads to the emergence of a transverse momentum scale $Q_{\text{med}}$ in the medium, which is apparent when multiple interactions are resummed. If this scale is well separated from the medium scale $T$, it can lead to novel logarithmic corrections as noted in Refs.~\cite{Liou:2013qya,Mehtar-Tani:2013pia,Iancu:2014kga,Blaizot:2014bha,Arnold:2021pin}. We leave an analysis of these aspects within the EFT-based approach developed here for future work.

\textit{ The jet function at one loop.} As a consistency check of our framework, we compute the one-loop result of ${\bf F}_{1,q}^{(1)}$ with a single medium interaction. The tree-level result vanishes since the Glauber gluons do not contribute directly to energy loss at leading power, while the one loop result is reads
\begin{align}\label{eq:F1q}
&{\bf F}_{1,q}^{(1)}(\epsilon,{\bfk},x^-,\mu)=\frac{\alpha_s(N_c^2-1)}{4\pi^2}\mu^{2\epsilon} \int \frac{{\rm d}^{2-2\epsilon}\bfq}{\bfk^2 \bfq^2} \frac{2 {\bfk} \cdot {\bfq}}{({\bfq}+{\bfk})^2}  \nonumber \\
& \int\frac{{\rm d}q^-}{q^-}\Theta\left(|\bfq|- \frac{q^-R}{2}\right)\Big[\delta(\epsilon)-\delta(q^--\epsilon)\Big] \nonumber \\
&\times\left\{1- \cos \left[\frac{({\bfq}+{\bfk})^2}{2q^-}x^-\right]\right\}\,,
\end{align}
which agrees with GLV results in Refs.~\cite{Gyulassy:2000fs,Gyulassy:2000er,Wiedemann:2000za}. The BFKL and threshold logarithms that lead to the anomalous dimensions will only appear at ${\cal O}(\alpha_s^2)$.
An alternative approach known as the Background Field Method based on the light-cone gauge $A^-=0$, which has been discussed extensively in the literature  yields identical results at leading order in perturbation theory. We discuss this in Appendix \ref{app:BFM}.

{\it Conclusions and Outlook.} 
In this letter, we derived for the first time an all-order factorization formula for inclusive jets in heavy-ion collisions in the framework of SCET for the phenomenologically relevant kinematic regime. The factorization leads to a series expansion involving an increasing number of subjet functions that encode the medium evolution of the jet, convolved with perturbative matching coefficients. We further write each subjet function as a series in the number of jet-medium interactions, while retaining radiative corrections to all orders. Each term in this series can be factorized in terms of a medium function that captures the universal microscopic physics of the medium and a process-dependent jet function, both of which have the same virtuality. 
The virtuality of the subjet functions is set by an emergent medium scale $Q_{\text{med}}$. The possibility that this scale is perturbative even in a strongly coupled medium when multiple scatterings are taken into account can precipitate further factorizations. This will determine the universality of the nonperturbative physics across distinct jet observables and will be an important question for future work. This work also opens up an avenue to address several other relevant questions including the matching and renormalization for the resolved case $\theta_c \ll R$, establishing a connection to the BDMPS-Z framework, the possibility of non-linear evolution in a dense medium, and utilizing the power of the EFT approach to compute higher orders in perturbation theory. In addition, we hope that this letter will serve as a starting point for a systematically improvable phenomenological analysis for jet substructure observables in heavy-ion collisions.

\section*{Acknowledgements}
V.V. and B.S.  are supported by startup funds from the University of South Dakota. Y. M.-T. is supported by the U.S. Department of Energy under Contract No. DE-SC0012704 and by Laboratory Directed Research and Development (LDRD) funds from Brookhaven Science Associates. FR was supported by the DOE, Office of Science, Contract No.~DE-AC05-06OR23177, under which Jefferson Science Associates, LLC operates Jefferson Lab, and in part by the DOE, Office of Science, Office of Nuclear Physics, Early Career Program under contract No. DE-SC0024358. 

\section*{Appendix}

\section{Operator definitions in SCET}
\label{app:Ocs}

While the operators and Hamiltonians used in the main text can be found in Ref.~\cite{Rothstein:2016bsq}, we collect them here for convenience. All of the SCET operators are defined in terms of gauge invariant building blocks
\small
\begin{align}
 \chi_{n} &= W_{n}^{\dagger}\xi_{n}, 
  \   \   \   \    
  W_{n} = \text{FT} \  {\bf P} \exp \Big\{ig\int_{-\infty}^0 {\rm d}s \, \bar{n}\cdot A_{n}(x+\bar{n}s)\Big\}  ,
  \nonumber\\
 \chi_{\rm s} &= S_{n}^{\dagger}\xi_{\rm s},  
   \   \   \   \    
   S_{n} = \text{FT} \ {\bf P} \exp \Big\{ig\int_{-\infty}^0 {\rm d}s\, n\cdot A_{\rm s}(x+s n)\Big\}
   , \nonumber\\
 & \mathcal{B}_{n \perp}^{\mu }\equiv \mathcal{B}_{n \perp}^{\mu a}t^a  = \frac{1}{g}\Big[W_{n}^{\dagger}iD_{n \perp}^{\mu}W_{ n}\Big], \nonumber \\
 \  \  
 &\mathcal{B}_{\rm s\perp}^{\mu }\equiv \mathcal{B}_{\rm s\perp}^{\mu a}t^a  = \frac{1}{g}\Big[S_{n}^{\dagger}iD_{\rm s \perp}^{\mu} S_{n}\Big] 
 .
\end{align}
\normalsize
Here, FT denotes the Fourier transform. These operators encode bare quarks and gluons dressed by Wilson lines.

Analogous to the quark jet function introduced for the hard, hc factorization, we can also define the gluon jet function as
\begin{align}
&\!J_g(z,\omega_J,\mu)\!=\!\frac{\omega}{4C_FN_c}\int\!\! {\rm d}s\, e^{is\omega}\tr\left[\mathcal{B}_{n\perp\mu}(0)\rho_M\mathcal{B}_{n\perp}^{\mu}(s \bar n)\mathcal{M} \right]. 
\label{eq:gjet}
\end{align}
Further, the subjet function $\mathcal{S}_1$ in the hc, collinear-soft stage of the factorization is evaluated with the Hamiltonian 
\small
\begin{equation}
\label{eq:H2}
\int\! {\rm d}t\, H(t)\!=\!\int {\rm d}t \left(H_{\rm cs}(t)+H_{\rm s}(t)+ H_{\rm cs\text{-}s}(t)\right) + \int {\rm d} s\, \mathbf{O}_{\rm cs\text{-}s}(sn) \,.
\end{equation}
 Here, $H_{\text{cs}}$ is the standard collinear SCET Hamiltonian, and $H_{\text{s}}$ describes the dynamics of the soft field, which is the full QCD Hamiltonian. Moreover, $H_{\text{cs-s}}$ describes the forward scattering of the collinear-soft gluon off a soft medium parton. 

The Hamiltonian interaction density of the collinear-soft mode with the soft field is given by 
\begin{align}
\label{EFTOp}
\mathcal{H}_{\rm cs\text{-}s}
 &= C_G\frac{i}{2}f^{abc}\mathcal{B}_{n \perp\mu}^b\frac{\bar n}{2}\cdot(\mathcal{P}+\mathcal{P}^{\dagger})\mathcal{B}_{n \perp}^{c\mu}\frac{1}{\mathcal{P}_{\perp}^2}\mathcal{O}_{\rm s}^a
 \,. 
\end{align}
Here, $C_G(\mu)=8\pi\alpha_s(\mu)$ and $\mathcal{O}_{ \rm s}^a =\sum_{j \in \{q,\bar q, g\}}\mathcal{O}_{\rm s}^{j,a}$ with
\small
\begin{align}
\mathcal{O}_{\rm s}^{q, a} &= \bar{\chi}_{\rm s}t^a\frac{\slashed{n}}{2}\chi_{\rm s}, 
 \   \   \    
\mathcal{O}_{\rm s}^{g, a}=  \frac{i}{2}f^{abc}\mathcal{B}_{\rm s \perp\mu}^b\frac{n}{2}\cdot(\mathcal{P}+\mathcal{P}^{\dagger})\mathcal{B}_{\rm s \perp}^{c\mu}. 
 \label{eq:Opbarn}
 \end{align}
The medium-induced radiation to all orders is taken into account by the operator $\mathbf{O}_{\text{cs-s}}$, which is defined as 
\begin{align}\label{eq:Oc-s}
\mathbf{O}_{\text{cs-s}}(sn) = \int {\rm d}^2\bfq \frac{1}{\bfq^2}\Big[O^{ab}_{\text{cs}}\frac{1}{\mathcal{P}_{\perp}^2}\mathcal{O}_{\text{s}}^b\Big](sn,\bfq) t^a\,,
\end{align} 
where $O_{\text{cs}}$ and $O_{\text{s}}$ are gauge invariant operators built out of collinear-soft and soft fields, respectively. The operator in Eq.~(\ref{eq:Oc-s}) is given by
\begin{align}
&O_{\rm cs}^{ab} = 8\pi \alpha_s\Bigg[\mathcal{P}_{\perp}^{\mu}S_n^TW_{n}\mathcal{P}_{\perp \mu} -\mathcal{P}_{\mu}^{\perp}g \mathcal{\tilde B}_{\rm s\perp}^{n\mu}S_n^TW_n -\nonumber \\
& S_n^TW_ng \mathcal{\tilde B}^{n \mu}_{\perp}\mathcal{P}_{\mu}^{\perp}- g \mathcal{\tilde B}_{\rm s \perp}^{n\mu}S_n^TW_ng\mathcal{\tilde B}^{n}_{\perp \mu}-\frac{n_{\mu}\bar n_{\nu}}{2}S_n^T ig \tilde G^{\mu\nu}W_n\Bigg]^{ab},
\label{eq:LPT}
\end{align}
\normalsize
where $\tilde G^{\mu\nu}$ is the gluon field strength tensor.

\section{ Renormalization group consistency of factorization}
\label{app:RG}

Here, we discuss the anomalous dimensions for the various functions that appear in our factorization formulas. We start with Eq.~(\ref{eq:J-S-fact}). For a given function $f(z, \omega_J,\mu)$, the RG equation and the associated anomalous dimension are defined through 
 \beq 
\! \mu \frac{{\rm d}}{{\rm d}\mu} f^i(z,\omega_J,\mu) = \sum_j\!\int_z^1\! \frac{{\rm d}z'}{z'} \gamma^{ij}_{f}\left(\frac{z}{z'},\mu\right)f^j(z',\omega_J,\mu) 
 \eeq
with $i,j \in \{q,g\}$. For the function ${\cal C}_{q\to 1}$, we find
\begin{align}
\gamma^{qq}_{{\cal C}_{q\to 1}}(z) &= \delta(1-z) \frac{\alpha_s C_F}{2\pi}(4 \ell +3) -\frac{\alpha_sC_F}{\pi}(1+z)\nonumber \\
\gamma^{qg}_{{\cal C}_{q\to 1}}(z)&=\frac{\alpha_sC_F}{\pi}P_{gq}(z)\,,
\end{align}
with $\ell= \ln(\mu/(\omega_JR/2))$ and $P_{ij}$ are the DGLAP splitting functions. The anomalous dimension of the collinear-soft function ${\cal S}_1$ at one-loop reads
\begin{align}\label{eq:gammaS}
\gamma^q_{{\cal S}_1}(z)= -\delta(1-z)\frac{4\alpha_sC_F}{2\pi} \ell +\frac{\alpha_sC_F}{2\pi}\frac{4}{(1-z)}_+\,,
\end{align}
where the variable $z$ is defined as $\epsilon = (1-z) \omega_J'$. We find $\gamma^{q}_{{\cal S}_1}(z)+ \gamma^{qq}_{{\cal C}_{q\to 1}}(z) =\alpha_sC_F/\pi\, P_{qq}(z) $, as required by RG consistency in Eq.~(\ref{eq:J-S-fact}). The refactorization leads to the RG running between the scales $p_TR(1-z)$ and $p_TR$ resumming the leading threshold logarithms. A full threshold resummation would also require us to refactorize the hard function in Eq.~(\ref{eq:factI}). This has been carried out for the vacuum in Refs.~\cite{deFlorian:2007fv,Dai:2017dpc,Liu:2017pbb,Neill:2021std}, and we leave an extension to the medium case for future work.

The function $\varphi(\bfk)$ defined in Eq.~(\ref{eq:MedFn}) obeys the BFKL evolution equation in the rapidity renormalization scale $\nu$, and the evolution in $\mu$ is determined by the QCD  beta function
\begin{align}
\label{eq:BRG}
 \frac{{\rm d} \varphi(\bfk;\nu;\mu)}{{\rm d}\ln \nu} 
 = &\,\frac{\alpha_s N_c}{\pi^2} \int {\rm d}^2\bfu \Bigg[\frac{\varphi(\bfu;\nu;\mu)}{({\bfu}-{\bfk})^2} -\frac{\bfk^2\varphi(\bfk;\nu;\mu) }{2\bfu^2({\bfu}-{\bfk})^2}\Bigg]\,,\nonumber\\
 \frac{{\rm d}\varphi(\bfk;\nu;\mu)}{{\rm d}\ln \mu} = &-\frac{\alpha_s \beta_0}{\pi}\varphi(\bfk;\nu;\mu)\,.
\end{align}

\section{Background Field Method}
\label{app:BFM}

In this Appendix, we use the background field method as an alternative approach to computing the fixed order result in \eqn{eq:F1q}. 

In the background field method, the medium is represented entirely by a classical background Glauber (G) field that is characterized by a vanishing longitudinal component of the four-momentum, i.e. $k^-=0$.  To isolate such a mode, we split the cs field into a cs radiative part $p_{\rm cs}\sim p_TR(1, R^2, R)$ and an off-shell Glauber mode  $p_{\rm G}\sim p_TR(R, R^2, R)$  that mediates $t$-channel gluon exchange with the medium~\cite{Balitsky:1995ub,Mehtar-Tani:2006vpj,Mehtar-Tani:2010ebp,Blaizot:2015lma}, $A_{\rm cs} \to A_{\rm cs} + A_{\rm G}$. Here, we work in the light-cone gauge $\bar n \cdot A=A^-=0$  as is customary in high-energy scattering processes. Only the plus component of the G field contributes to leading power in $R$ due to the fact that in any given collinear direction $i$, in the $A_{\rm cs}$ Wilson line in the function $\mathcal{S}_m$, Eq.~(\ref{eq:soft-funct}), $n_i\cdot A_\Gl \sim n_i^-  A^+_\Gl- n_{i,\perp} \cdot  A_{\Gl,\perp}  \sim A^+_\Gl$. Here, we used that $A^+_\Gl-\sim A_{\Gl,\perp} $ for a dynamical plasma and $n_{i,\perp} \sim R \ll n_i^- \sim 1$. The corresponding Hamiltonian density reads 
\begin{align}
 \cH_{\rm int}[A_\cs,A_\Gl ] &= 2ig \tr (\del^- A_{\cs,\perp}^\mu ) [A_\Gl^+,A_{\cs,\mu\perp}] \nonumber\\ 
 &= -g (\del^- A_{\cs,\perp}^{a,\mu}) A_\Gl^{b,+} A^c_{\cs, \mu\perp} f^{abc}\,.
\end{align}
The remaining parts of the Hamiltonian encompass the full QCD dynamics for the cs and G fields, which reside on different sides of the rapidity cut. However, to perform explicit calculations of observables involving the expectation values of the G field, additional assumptions are needed. These include how the G field is sourced by the plasma's color charges and the extent of its correlation length. For fixed order calculation one invokes that the G field can be described by a classical field that obeys the Yang-Mills equations of motion \cite{Mehtar-Tani:2006vpj}, $ A^+_\Gl (x) = -\del^2_\perp A^+_\Gl (x^-,\bfx,0^+)= J^+(x^-,\bfx,0^+)$, in the center of mass frame. Second, $A_G$ is typically treated as a stochastic field obeying Gaussian statistics, cf. \eqn{eq:Gfield-corr}. This procedure is reminiscent of the high-energy factorization approach in the Color Glass Condensate Effective Theory~\cite{Gelis:2010nm}. 

The single subjet Wilson line $U_1(n)$, Eq.~(\ref{eq:CSWilson}), can be expressed as a sum over the number of cs and Glauber modes, $n$ and $m$, respectively: 
\beq 
U_1\equiv \sum_{n=0,m=0}^\infty U_1^{(n,m)} \,. 
\eeq
Here, we shall simply present the next-to-leading order contribution in both the cs and Glauber fields. We want to compute the one-gluon production amplitude $\langle k | U |0\rangle $. 
 
For a large medium length, we recover the sum of the Lipatov vertex in light-cone gauge and the early radiation of a soft gluon followed by the rescattering of the fast jet color charge \cite{Wiedemann:2000za,Gyulassy:2000fs} (for details see Appendix A in \cite{Mehtar-Tani:2011lic}) 
\begin{align}\label{eq:WL-opac1-3}
\langle q | U |0\rangle \to & - 2g^2 f^{abc}t^c \int_{\bfk} \left[\frac{\epsilon^a_{\lambda}\cdot (\bfq-\bfk)}{ (\bfq-\bfk)^2} -\frac{\epsilon^a_{\lambda}\cdot \bfq}{ \bfq^2}\right] \nonumber \\ &\times\int_0^{+\infty} \rmd y^- \rme^{i\frac{\bfq^2}{2q^-} y^-}A^{+,b}_\Gl(y^-,\bfk) \,\nonumber \\ 
&  -2g t^a \frac{\epsilon^a_{\lambda}(q)\cdot \bfq}{ \bfq^2} 
\int_0^{+\infty} \rmd y^-  \, (ig) t^b A^{+,b}_\Gl(y^-) \,.
\end{align}
Squaring the amplitude, accounting for virtual corrections, and integrating over the momentum of the radiated gluons while incorporating the measurement function, we obtain
\begin{align}\label{eq:FctA}
 {\cal S}_{1}(\epsilon,\mu) &\simeq \delta(\epsilon) + \frac{4 g^2 C_F}{2(2\pi)}\mu^{-\epsilon}\int \frac{\rmd q^-}{q^-} \int \frac{\rmd^{2+\epsilon}  \bfq}{(2\pi)^{2+\epsilon} \bfq^2} (2\pi)\nonumber\\
 & \times\left[\delta(q^--\epsilon)-\delta(\epsilon)\right] \Theta\left( |\bfq|-\frac{R q^-}{2}\right)\nonumber\\
&+\int_{0}^\infty{\rm d}x^- \int \frac{{\rm d}^2\bfk}{(2\pi)^3} {\bf F}_{q,1}(\epsilon, \mu; \bfk,x^-) \, \varphi_0(\bfk,x^-)\,, \nonumber \\
\end{align} 
with ${\bf F}_{q,1}$ given in \eqn{eq:F1q}. Integrating over $\bfq$ and $q^-$ in the second term yields the integral over $\mu$ of the cs anomalous dimension given by $\gamma_{\cS_1}$, \eqn{eq:gammaS}. 

Here $\varphi_0(\bfk)=\varphi(\bfk,\nu=0)$ is the initial condition for the BFKL evolution, which, to leading order, reads $\varphi_0(\bfk) \equiv g^2/\bfk^2$ for $|\bfk| \gg m_D$ and is related to the correlator of background fields as follows \cite{Blaizot:2015lma}
\small
\begin{align}\label{eq:Gfield-corr}
&\text{Tr}\Big[A_\Gl^+(\bfk,x^-)\rho_M A^{+,\ast}_\Gl(-\bfk',x'^-) \Big]\equiv \langle A^a_\Gl(\bfk,x^-) A^{\ast,a}_\Gl(-\bfk',x'^-) \rangle \nonumber \\
&=  \frac{1}{\bfk^2}\, \varphi_0(\bfk,x^-) \delta(x^--x'^-) (2\pi)^2\delta(\bfk-\bfk')  \,.
\end{align}
\normalsize

\bibliography{SemiJet.bib}

\begin{thebibliography}{94}%
\makeatletter
\providecommand \@ifxundefined [1]{%
 \@ifx{#1\undefined}
}%
\providecommand \@ifnum [1]{%
 \ifnum #1\expandafter \@firstoftwo
 \else \expandafter \@secondoftwo
 \fi
}%
\providecommand \@ifx [1]{%
 \ifx #1\expandafter \@firstoftwo
 \else \expandafter \@secondoftwo
 \fi
}%
\providecommand \natexlab [1]{#1}%
\providecommand \enquote  [1]{``#1''}%
\providecommand \bibnamefont  [1]{#1}%
\providecommand \bibfnamefont [1]{#1}%
\providecommand \citenamefont [1]{#1}%
\providecommand \href@noop [0]{\@secondoftwo}%
\providecommand \href [0]{\begingroup \@sanitize@url \@href}%
\providecommand \@href[1]{\@@startlink{#1}\@@href}%
\providecommand \@@href[1]{\endgroup#1\@@endlink}%
\providecommand \@sanitize@url [0]{\catcode `\\12\catcode `\$12\catcode
  `\&12\catcode `\#12\catcode `\^12\catcode `\_12\catcode `\%12\relax}%
\providecommand \@@startlink[1]{}%
\providecommand \@@endlink[0]{}%
\providecommand \url  [0]{\begingroup\@sanitize@url \@url }%
\providecommand \@url [1]{\endgroup\@href {#1}{\urlprefix }}%
\providecommand \urlprefix  [0]{URL }%
\providecommand \Eprint [0]{\href }%
\providecommand \doibase [0]{http://dx.doi.org/}%
\providecommand \selectlanguage [0]{\@gobble}%
\providecommand \bibinfo  [0]{\@secondoftwo}%
\providecommand \bibfield  [0]{\@secondoftwo}%
\providecommand \translation [1]{[#1]}%
\providecommand \BibitemOpen [0]{}%
\providecommand \bibitemStop [0]{}%
\providecommand \bibitemNoStop [0]{.\EOS\space}%
\providecommand \EOS [0]{\spacefactor3000\relax}%
\providecommand \BibitemShut  [1]{\csname bibitem#1\endcsname}%
\let\auto@bib@innerbib\@empty
\bibitem [{\citenamefont {Bjorken}(1982)}]{Bjorken:1982tu}%
  \BibitemOpen
  \bibfield  {author} {\bibinfo {author} {\bibfnamefont {J.~D.}\ \bibnamefont
  {Bjorken}},\ }\href {\doibase FERMILAB-PUB-82-059-THY} {\  (\bibinfo {year}
  {1982}),\ FERMILAB-PUB-82-059-THY}\BibitemShut {NoStop}%
\bibitem [{\citenamefont {Arsene}\ \emph {et~al.}(2005)\citenamefont {Arsene}
  \emph {et~al.}}]{BRAHMS:2004adc}%
  \BibitemOpen
  \bibfield  {author} {\bibinfo {author} {\bibfnamefont {I.}~\bibnamefont
  {Arsene}} \emph {et~al.} (\bibinfo {collaboration} {BRAHMS}),\ }\href
  {\doibase 10.1016/j.nuclphysa.2005.02.130} {\bibfield  {journal} {\bibinfo
  {journal} {Nucl. Phys. A}\ }\textbf {\bibinfo {volume} {757}},\ \bibinfo
  {pages} {1} (\bibinfo {year} {2005})},\ \Eprint
  {http://arxiv.org/abs/nucl-ex/0410020} {arXiv:nucl-ex/0410020} \BibitemShut
  {NoStop}%
\bibitem [{\citenamefont {Back}\ \emph {et~al.}(2005)\citenamefont {Back} \emph
  {et~al.}}]{PHOBOS:2004zne}%
  \BibitemOpen
  \bibfield  {author} {\bibinfo {author} {\bibfnamefont {B.~B.}\ \bibnamefont
  {Back}} \emph {et~al.} (\bibinfo {collaboration} {PHOBOS}),\ }\href {\doibase
  10.1016/j.nuclphysa.2005.03.084} {\bibfield  {journal} {\bibinfo  {journal}
  {Nucl. Phys. A}\ }\textbf {\bibinfo {volume} {757}},\ \bibinfo {pages} {28}
  (\bibinfo {year} {2005})},\ \Eprint {http://arxiv.org/abs/nucl-ex/0410022}
  {arXiv:nucl-ex/0410022} \BibitemShut {NoStop}%
\bibitem [{\citenamefont {Adams}\ \emph {et~al.}(2005)\citenamefont {Adams}
  \emph {et~al.}}]{STAR:2005gfr}%
  \BibitemOpen
  \bibfield  {author} {\bibinfo {author} {\bibfnamefont {J.}~\bibnamefont
  {Adams}} \emph {et~al.} (\bibinfo {collaboration} {STAR}),\ }\href {\doibase
  10.1016/j.nuclphysa.2005.03.085} {\bibfield  {journal} {\bibinfo  {journal}
  {Nucl. Phys. A}\ }\textbf {\bibinfo {volume} {757}},\ \bibinfo {pages} {102}
  (\bibinfo {year} {2005})},\ \Eprint {http://arxiv.org/abs/nucl-ex/0501009}
  {arXiv:nucl-ex/0501009} \BibitemShut {NoStop}%
\bibitem [{\citenamefont {Adcox}\ \emph {et~al.}(2005)\citenamefont {Adcox}
  \emph {et~al.}}]{PHENIX:2004vcz}%
  \BibitemOpen
  \bibfield  {author} {\bibinfo {author} {\bibfnamefont {K.}~\bibnamefont
  {Adcox}} \emph {et~al.} (\bibinfo {collaboration} {PHENIX}),\ }\href
  {\doibase 10.1016/j.nuclphysa.2005.03.086} {\bibfield  {journal} {\bibinfo
  {journal} {Nucl. Phys. A}\ }\textbf {\bibinfo {volume} {757}},\ \bibinfo
  {pages} {184} (\bibinfo {year} {2005})},\ \Eprint
  {http://arxiv.org/abs/nucl-ex/0410003} {arXiv:nucl-ex/0410003} \BibitemShut
  {NoStop}%
\bibitem [{\citenamefont {Aad}\ \emph {et~al.}(2010)\citenamefont {Aad} \emph
  {et~al.}}]{ATLAS:2010isq}%
  \BibitemOpen
  \bibfield  {author} {\bibinfo {author} {\bibfnamefont {G.}~\bibnamefont
  {Aad}} \emph {et~al.} (\bibinfo {collaboration} {ATLAS}),\ }\href {\doibase
  10.1103/PhysRevLett.105.252303} {\bibfield  {journal} {\bibinfo  {journal}
  {Phys. Rev. Lett.}\ }\textbf {\bibinfo {volume} {105}},\ \bibinfo {pages}
  {252303} (\bibinfo {year} {2010})},\ \Eprint {http://arxiv.org/abs/1011.6182}
  {arXiv:1011.6182 [hep-ex]} \BibitemShut {NoStop}%
\bibitem [{\citenamefont {Aamodt}\ \emph {et~al.}(2011)\citenamefont {Aamodt}
  \emph {et~al.}}]{ALICE:2010yje}%
  \BibitemOpen
  \bibfield  {author} {\bibinfo {author} {\bibfnamefont {K.}~\bibnamefont
  {Aamodt}} \emph {et~al.} (\bibinfo {collaboration} {ALICE}),\ }\href
  {\doibase 10.1016/j.physletb.2010.12.020} {\bibfield  {journal} {\bibinfo
  {journal} {Phys. Lett. B}\ }\textbf {\bibinfo {volume} {696}},\ \bibinfo
  {pages} {30} (\bibinfo {year} {2011})},\ \Eprint
  {http://arxiv.org/abs/1012.1004} {arXiv:1012.1004 [nucl-ex]} \BibitemShut
  {NoStop}%
\bibitem [{\citenamefont {Chatrchyan}\ \emph {et~al.}(2011)\citenamefont
  {Chatrchyan} \emph {et~al.}}]{CMS:2011iwn}%
  \BibitemOpen
  \bibfield  {author} {\bibinfo {author} {\bibfnamefont {S.}~\bibnamefont
  {Chatrchyan}} \emph {et~al.} (\bibinfo {collaboration} {CMS}),\ }\href
  {\doibase 10.1103/PhysRevC.84.024906} {\bibfield  {journal} {\bibinfo
  {journal} {Phys. Rev. C}\ }\textbf {\bibinfo {volume} {84}},\ \bibinfo
  {pages} {024906} (\bibinfo {year} {2011})},\ \Eprint
  {http://arxiv.org/abs/1102.1957} {arXiv:1102.1957 [nucl-ex]} \BibitemShut
  {NoStop}%
\bibitem [{\citenamefont {Aaboud}\ \emph {et~al.}(2019)\citenamefont {Aaboud}
  \emph {et~al.}}]{ATLAS:2018gwx}%
  \BibitemOpen
  \bibfield  {author} {\bibinfo {author} {\bibfnamefont {M.}~\bibnamefont
  {Aaboud}} \emph {et~al.} (\bibinfo {collaboration} {ATLAS}),\ }\href
  {\doibase 10.1016/j.physletb.2018.10.076} {\bibfield  {journal} {\bibinfo
  {journal} {Phys. Lett. B}\ }\textbf {\bibinfo {volume} {790}},\ \bibinfo
  {pages} {108} (\bibinfo {year} {2019})},\ \Eprint
  {http://arxiv.org/abs/1805.05635} {arXiv:1805.05635 [nucl-ex]} \BibitemShut
  {NoStop}%
\bibitem [{\citenamefont {Sirunyan}\ \emph {et~al.}(2021)\citenamefont
  {Sirunyan} \emph {et~al.}}]{CMS:2021vui}%
  \BibitemOpen
  \bibfield  {author} {\bibinfo {author} {\bibfnamefont {A.~M.}\ \bibnamefont
  {Sirunyan}} \emph {et~al.} (\bibinfo {collaboration} {CMS}),\ }\href
  {\doibase 10.1007/JHEP05(2021)284} {\bibfield  {journal} {\bibinfo  {journal}
  {JHEP}\ }\textbf {\bibinfo {volume} {05}},\ \bibinfo {pages} {284} (\bibinfo
  {year} {2021})},\ \Eprint {http://arxiv.org/abs/2102.13080} {arXiv:2102.13080
  [hep-ex]} \BibitemShut {NoStop}%
\bibitem [{\citenamefont {Acharya}\ \emph {et~al.}(2024)\citenamefont {Acharya}
  \emph {et~al.}}]{ALICE:2023waz}%
  \BibitemOpen
  \bibfield  {author} {\bibinfo {author} {\bibfnamefont {S.}~\bibnamefont
  {Acharya}} \emph {et~al.} (\bibinfo {collaboration} {ALICE}),\ }\href
  {\doibase 10.1016/j.physletb.2023.138412} {\bibfield  {journal} {\bibinfo
  {journal} {Phys. Lett. B}\ }\textbf {\bibinfo {volume} {849}},\ \bibinfo
  {pages} {138412} (\bibinfo {year} {2024})},\ \Eprint
  {http://arxiv.org/abs/2303.00592} {arXiv:2303.00592 [nucl-ex]} \BibitemShut
  {NoStop}%
\bibitem [{\citenamefont {Connors}\ \emph {et~al.}(2018)\citenamefont
  {Connors}, \citenamefont {Nattrass}, \citenamefont {Reed},\ and\
  \citenamefont {Salur}}]{Connors:2017ptx}%
  \BibitemOpen
  \bibfield  {author} {\bibinfo {author} {\bibfnamefont {M.}~\bibnamefont
  {Connors}}, \bibinfo {author} {\bibfnamefont {C.}~\bibnamefont {Nattrass}},
  \bibinfo {author} {\bibfnamefont {R.}~\bibnamefont {Reed}}, \ and\ \bibinfo
  {author} {\bibfnamefont {S.}~\bibnamefont {Salur}},\ }\href {\doibase
  10.1103/RevModPhys.90.025005} {\bibfield  {journal} {\bibinfo  {journal}
  {Rev. Mod. Phys.}\ }\textbf {\bibinfo {volume} {90}},\ \bibinfo {pages}
  {025005} (\bibinfo {year} {2018})},\ \Eprint
  {http://arxiv.org/abs/1705.01974} {arXiv:1705.01974 [nucl-ex]} \BibitemShut
  {NoStop}%
\bibitem [{\citenamefont {Currie}\ \emph {et~al.}(2017)\citenamefont {Currie},
  \citenamefont {Glover},\ and\ \citenamefont {Pires}}]{Currie:2016bfm}%
  \BibitemOpen
  \bibfield  {author} {\bibinfo {author} {\bibfnamefont {J.}~\bibnamefont
  {Currie}}, \bibinfo {author} {\bibfnamefont {E.~W.~N.}\ \bibnamefont
  {Glover}}, \ and\ \bibinfo {author} {\bibfnamefont {J.}~\bibnamefont
  {Pires}},\ }\href {\doibase 10.1103/PhysRevLett.118.072002} {\bibfield
  {journal} {\bibinfo  {journal} {Phys. Rev. Lett.}\ }\textbf {\bibinfo
  {volume} {118}},\ \bibinfo {pages} {072002} (\bibinfo {year} {2017})},\
  \Eprint {http://arxiv.org/abs/1611.01460} {arXiv:1611.01460 [hep-ph]}
  \BibitemShut {NoStop}%
\bibitem [{\citenamefont {Czakon}\ \emph {et~al.}(2019)\citenamefont {Czakon},
  \citenamefont {van Hameren}, \citenamefont {Mitov},\ and\ \citenamefont
  {Poncelet}}]{Czakon:2019tmo}%
  \BibitemOpen
  \bibfield  {author} {\bibinfo {author} {\bibfnamefont {M.}~\bibnamefont
  {Czakon}}, \bibinfo {author} {\bibfnamefont {A.}~\bibnamefont {van Hameren}},
  \bibinfo {author} {\bibfnamefont {A.}~\bibnamefont {Mitov}}, \ and\ \bibinfo
  {author} {\bibfnamefont {R.}~\bibnamefont {Poncelet}},\ }\href {\doibase
  10.1007/JHEP10(2019)262} {\bibfield  {journal} {\bibinfo  {journal} {JHEP}\
  }\textbf {\bibinfo {volume} {10}},\ \bibinfo {pages} {262} (\bibinfo {year}
  {2019})},\ \Eprint {http://arxiv.org/abs/1907.12911} {arXiv:1907.12911
  [hep-ph]} \BibitemShut {NoStop}%
\bibitem [{\citenamefont {Larkoski}\ \emph {et~al.}(2020)\citenamefont
  {Larkoski}, \citenamefont {Moult},\ and\ \citenamefont
  {Nachman}}]{Larkoski:2017jix}%
  \BibitemOpen
  \bibfield  {author} {\bibinfo {author} {\bibfnamefont {A.~J.}\ \bibnamefont
  {Larkoski}}, \bibinfo {author} {\bibfnamefont {I.}~\bibnamefont {Moult}}, \
  and\ \bibinfo {author} {\bibfnamefont {B.}~\bibnamefont {Nachman}},\ }\href
  {\doibase 10.1016/j.physrep.2019.11.001} {\bibfield  {journal} {\bibinfo
  {journal} {Phys. Rept.}\ }\textbf {\bibinfo {volume} {841}},\ \bibinfo
  {pages} {1} (\bibinfo {year} {2020})},\ \Eprint
  {http://arxiv.org/abs/1709.04464} {arXiv:1709.04464 [hep-ph]} \BibitemShut
  {NoStop}%
\bibitem [{\citenamefont {Kogler}\ \emph {et~al.}(2019)\citenamefont {Kogler}
  \emph {et~al.}}]{Asquith:2018igt}%
  \BibitemOpen
  \bibfield  {author} {\bibinfo {author} {\bibfnamefont {R.}~\bibnamefont
  {Kogler}} \emph {et~al.},\ }\href {\doibase 10.1103/RevModPhys.91.045003}
  {\bibfield  {journal} {\bibinfo  {journal} {Rev. Mod. Phys.}\ }\textbf
  {\bibinfo {volume} {91}},\ \bibinfo {pages} {045003} (\bibinfo {year}
  {2019})},\ \Eprint {http://arxiv.org/abs/1803.06991} {arXiv:1803.06991
  [hep-ex]} \BibitemShut {NoStop}%
\bibitem [{\citenamefont {Marzani}\ \emph {et~al.}(2019)\citenamefont
  {Marzani}, \citenamefont {Soyez},\ and\ \citenamefont
  {Spannowsky}}]{Marzani:2019hun}%
  \BibitemOpen
  \bibfield  {author} {\bibinfo {author} {\bibfnamefont {S.}~\bibnamefont
  {Marzani}}, \bibinfo {author} {\bibfnamefont {G.}~\bibnamefont {Soyez}}, \
  and\ \bibinfo {author} {\bibfnamefont {M.}~\bibnamefont {Spannowsky}},\
  }\href {\doibase 10.1007/978-3-030-15709-8} {\emph {\bibinfo {title}
  {{Looking inside jets: an introduction to jet substructure and boosted-object
  phenomenology}}}},\ Vol.\ \bibinfo {volume} {958}\ (\bibinfo  {publisher}
  {Springer},\ \bibinfo {year} {2019})\ \Eprint
  {http://arxiv.org/abs/1901.10342} {arXiv:1901.10342 [hep-ph]} \BibitemShut
  {NoStop}%
\bibitem [{\citenamefont {Bauer}\ \emph
  {et~al.}(2003{\natexlab{a}})\citenamefont {Bauer}, \citenamefont {Pirjol},\
  and\ \citenamefont {Stewart}}]{Bauer:2002aj}%
  \BibitemOpen
  \bibfield  {author} {\bibinfo {author} {\bibfnamefont {C.~W.}\ \bibnamefont
  {Bauer}}, \bibinfo {author} {\bibfnamefont {D.}~\bibnamefont {Pirjol}}, \
  and\ \bibinfo {author} {\bibfnamefont {I.~W.}\ \bibnamefont {Stewart}},\
  }\href@noop {} {\bibfield  {journal} {\bibinfo  {journal} {Phys. Rev. D}\
  }\textbf {\bibinfo {volume} {67}},\ \bibinfo {pages} {071502} (\bibinfo
  {year} {2003}{\natexlab{a}})},\ \Eprint {http://arxiv.org/abs/hep-ph/0211069}
  {arXiv:hep-ph/0211069} \BibitemShut {NoStop}%
\bibitem [{\citenamefont {Bauer}\ \emph
  {et~al.}(2003{\natexlab{b}})\citenamefont {Bauer}, \citenamefont {Pirjol},\
  and\ \citenamefont {Stewart}}]{Bauer:2003mga}%
  \BibitemOpen
  \bibfield  {author} {\bibinfo {author} {\bibfnamefont {C.~W.}\ \bibnamefont
  {Bauer}}, \bibinfo {author} {\bibfnamefont {D.}~\bibnamefont {Pirjol}}, \
  and\ \bibinfo {author} {\bibfnamefont {I.~W.}\ \bibnamefont {Stewart}},\
  }\href@noop {} {\bibfield  {journal} {\bibinfo  {journal} {Phys. Rev. D}\
  }\textbf {\bibinfo {volume} {68}},\ \bibinfo {pages} {034021} (\bibinfo
  {year} {2003}{\natexlab{b}})},\ \Eprint {http://arxiv.org/abs/hep-ph/0303156}
  {arXiv:hep-ph/0303156} \BibitemShut {NoStop}%
\bibitem [{\citenamefont {Bauer}\ \emph {et~al.}(2001)\citenamefont {Bauer},
  \citenamefont {Fleming}, \citenamefont {Pirjol},\ and\ \citenamefont
  {Stewart}}]{Bauer:2000yr}%
  \BibitemOpen
  \bibfield  {author} {\bibinfo {author} {\bibfnamefont {C.~W.}\ \bibnamefont
  {Bauer}}, \bibinfo {author} {\bibfnamefont {S.}~\bibnamefont {Fleming}},
  \bibinfo {author} {\bibfnamefont {D.}~\bibnamefont {Pirjol}}, \ and\ \bibinfo
  {author} {\bibfnamefont {I.~W.}\ \bibnamefont {Stewart}},\ }\href@noop {}
  {\bibfield  {journal} {\bibinfo  {journal} {Phys. Rev. D}\ }\textbf {\bibinfo
  {volume} {63}},\ \bibinfo {pages} {114020} (\bibinfo {year} {2001})},\
  \Eprint {http://arxiv.org/abs/hep-ph/0011336} {arXiv:hep-ph/0011336}
  \BibitemShut {NoStop}%
\bibitem [{\citenamefont {Bauer}\ and\ \citenamefont
  {Stewart}(2001)}]{Bauer:2001ct}%
  \BibitemOpen
  \bibfield  {author} {\bibinfo {author} {\bibfnamefont {C.~W.}\ \bibnamefont
  {Bauer}}\ and\ \bibinfo {author} {\bibfnamefont {I.~W.}\ \bibnamefont
  {Stewart}},\ }\href@noop {} {\bibfield  {journal} {\bibinfo  {journal} {Phys.
  Lett. B}\ }\textbf {\bibinfo {volume} {516}},\ \bibinfo {pages} {134}
  (\bibinfo {year} {2001})},\ \Eprint {http://arxiv.org/abs/hep-ph/0107001}
  {arXiv:hep-ph/0107001} \BibitemShut {NoStop}%
\bibitem [{\citenamefont {Bauer}\ \emph {et~al.}(2002)\citenamefont {Bauer},
  \citenamefont {Fleming}, \citenamefont {Pirjol}, \citenamefont {Rothstein},\
  and\ \citenamefont {Stewart}}]{Bauer:2002nz}%
  \BibitemOpen
  \bibfield  {author} {\bibinfo {author} {\bibfnamefont {C.~W.}\ \bibnamefont
  {Bauer}}, \bibinfo {author} {\bibfnamefont {S.}~\bibnamefont {Fleming}},
  \bibinfo {author} {\bibfnamefont {D.}~\bibnamefont {Pirjol}}, \bibinfo
  {author} {\bibfnamefont {I.~Z.}\ \bibnamefont {Rothstein}}, \ and\ \bibinfo
  {author} {\bibfnamefont {I.~W.}\ \bibnamefont {Stewart}},\ }\href@noop {}
  {\bibfield  {journal} {\bibinfo  {journal} {Phys. Rev. D}\ }\textbf {\bibinfo
  {volume} {66}},\ \bibinfo {pages} {014017} (\bibinfo {year} {2002})},\
  \Eprint {http://arxiv.org/abs/hep-ph/0202088} {arXiv:hep-ph/0202088}
  \BibitemShut {NoStop}%
\bibitem [{\citenamefont {Collins}\ \emph {et~al.}(1989)\citenamefont
  {Collins}, \citenamefont {Soper},\ and\ \citenamefont
  {Sterman}}]{Collins:1989gx}%
  \BibitemOpen
  \bibfield  {author} {\bibinfo {author} {\bibfnamefont {J.~C.}\ \bibnamefont
  {Collins}}, \bibinfo {author} {\bibfnamefont {D.~E.}\ \bibnamefont {Soper}},
  \ and\ \bibinfo {author} {\bibfnamefont {G.~F.}\ \bibnamefont {Sterman}},\
  }\href {\doibase 10.1142/9789814503266_0001} {\bibfield  {journal} {\bibinfo
  {journal} {Adv. Ser. Direct. High Energy Phys.}\ }\textbf {\bibinfo {volume}
  {5}},\ \bibinfo {pages} {1} (\bibinfo {year} {1989})},\ \Eprint
  {http://arxiv.org/abs/hep-ph/0409313} {arXiv:hep-ph/0409313} \BibitemShut
  {NoStop}%
\bibitem [{\citenamefont {Korchemsky}\ and\ \citenamefont
  {Sterman}(1999)}]{Korchemsky:1999kt}%
  \BibitemOpen
  \bibfield  {author} {\bibinfo {author} {\bibfnamefont {G.~P.}\ \bibnamefont
  {Korchemsky}}\ and\ \bibinfo {author} {\bibfnamefont {G.~F.}\ \bibnamefont
  {Sterman}},\ }\href {\doibase 10.1016/S0550-3213(99)00308-9} {\bibfield
  {journal} {\bibinfo  {journal} {Nucl. Phys. B}\ }\textbf {\bibinfo {volume}
  {555}},\ \bibinfo {pages} {335} (\bibinfo {year} {1999})},\ \Eprint
  {http://arxiv.org/abs/hep-ph/9902341} {arXiv:hep-ph/9902341} \BibitemShut
  {NoStop}%
\bibitem [{\citenamefont {Lee}\ and\ \citenamefont
  {Sterman}(2007)}]{Lee:2006nr}%
  \BibitemOpen
  \bibfield  {author} {\bibinfo {author} {\bibfnamefont {C.}~\bibnamefont
  {Lee}}\ and\ \bibinfo {author} {\bibfnamefont {G.~F.}\ \bibnamefont
  {Sterman}},\ }\href {\doibase 10.1103/PhysRevD.75.014022} {\bibfield
  {journal} {\bibinfo  {journal} {Phys. Rev. D}\ }\textbf {\bibinfo {volume}
  {75}},\ \bibinfo {pages} {014022} (\bibinfo {year} {2007})},\ \Eprint
  {http://arxiv.org/abs/hep-ph/0611061} {arXiv:hep-ph/0611061} \BibitemShut
  {NoStop}%
\bibitem [{\citenamefont {Landau}\ and\ \citenamefont
  {Pomeranchuk}(1953)}]{Landau:1953um}%
  \BibitemOpen
  \bibfield  {author} {\bibinfo {author} {\bibfnamefont {L.~D.}\ \bibnamefont
  {Landau}}\ and\ \bibinfo {author} {\bibfnamefont {I.}~\bibnamefont
  {Pomeranchuk}},\ }\href@noop {} {\bibfield  {journal} {\bibinfo  {journal}
  {Dokl. Akad. Nauk Ser. Fiz.}\ }\textbf {\bibinfo {volume} {92}},\ \bibinfo
  {pages} {535} (\bibinfo {year} {1953})}\BibitemShut {NoStop}%
\bibitem [{\citenamefont {Migdal}(1956)}]{Migdal:1956tc}%
  \BibitemOpen
  \bibfield  {author} {\bibinfo {author} {\bibfnamefont {A.~B.}\ \bibnamefont
  {Migdal}},\ }\href {\doibase 10.1103/PhysRev.103.1811} {\bibfield  {journal}
  {\bibinfo  {journal} {Phys. Rev.}\ }\textbf {\bibinfo {volume} {103}},\
  \bibinfo {pages} {1811} (\bibinfo {year} {1956})}\BibitemShut {NoStop}%
\bibitem [{\citenamefont {Gyulassy}\ and\ \citenamefont
  {Wang}(1994)}]{Gyulassy:1993hr}%
  \BibitemOpen
  \bibfield  {author} {\bibinfo {author} {\bibfnamefont {M.}~\bibnamefont
  {Gyulassy}}\ and\ \bibinfo {author} {\bibfnamefont {X.-n.}\ \bibnamefont
  {Wang}},\ }\href {\doibase 10.1016/0550-3213(94)90079-5} {\bibfield
  {journal} {\bibinfo  {journal} {Nucl. Phys. B}\ }\textbf {\bibinfo {volume}
  {420}},\ \bibinfo {pages} {583} (\bibinfo {year} {1994})},\ \Eprint
  {http://arxiv.org/abs/nucl-th/9306003} {arXiv:nucl-th/9306003} \BibitemShut
  {NoStop}%
\bibitem [{\citenamefont {Wang}\ \emph {et~al.}(1995)\citenamefont {Wang},
  \citenamefont {Gyulassy},\ and\ \citenamefont {Plumer}}]{Wang:1994fx}%
  \BibitemOpen
  \bibfield  {author} {\bibinfo {author} {\bibfnamefont {X.-N.}\ \bibnamefont
  {Wang}}, \bibinfo {author} {\bibfnamefont {M.}~\bibnamefont {Gyulassy}}, \
  and\ \bibinfo {author} {\bibfnamefont {M.}~\bibnamefont {Plumer}},\ }\href
  {\doibase 10.1103/PhysRevD.51.3436} {\bibfield  {journal} {\bibinfo
  {journal} {Phys. Rev. D}\ }\textbf {\bibinfo {volume} {51}},\ \bibinfo
  {pages} {3436} (\bibinfo {year} {1995})},\ \Eprint
  {http://arxiv.org/abs/hep-ph/9408344} {arXiv:hep-ph/9408344} \BibitemShut
  {NoStop}%
\bibitem [{\citenamefont {Baier}\ \emph {et~al.}(1995)\citenamefont {Baier},
  \citenamefont {Dokshitzer}, \citenamefont {Peigne},\ and\ \citenamefont
  {Schiff}}]{Baier:1994bd}%
  \BibitemOpen
  \bibfield  {author} {\bibinfo {author} {\bibfnamefont {R.}~\bibnamefont
  {Baier}}, \bibinfo {author} {\bibfnamefont {Y.~L.}\ \bibnamefont
  {Dokshitzer}}, \bibinfo {author} {\bibfnamefont {S.}~\bibnamefont {Peigne}},
  \ and\ \bibinfo {author} {\bibfnamefont {D.}~\bibnamefont {Schiff}},\ }\href
  {\doibase 10.1016/0370-2693(94)01617-L} {\bibfield  {journal} {\bibinfo
  {journal} {Phys. Lett. B}\ }\textbf {\bibinfo {volume} {345}},\ \bibinfo
  {pages} {277} (\bibinfo {year} {1995})},\ \Eprint
  {http://arxiv.org/abs/hep-ph/9411409} {arXiv:hep-ph/9411409} \BibitemShut
  {NoStop}%
\bibitem [{\citenamefont {Baier}\ \emph
  {et~al.}(1997{\natexlab{a}})\citenamefont {Baier}, \citenamefont
  {Dokshitzer}, \citenamefont {Mueller}, \citenamefont {Peigne},\ and\
  \citenamefont {Schiff}}]{Baier:1996kr}%
  \BibitemOpen
  \bibfield  {author} {\bibinfo {author} {\bibfnamefont {R.}~\bibnamefont
  {Baier}}, \bibinfo {author} {\bibfnamefont {Y.~L.}\ \bibnamefont
  {Dokshitzer}}, \bibinfo {author} {\bibfnamefont {A.~H.}\ \bibnamefont
  {Mueller}}, \bibinfo {author} {\bibfnamefont {S.}~\bibnamefont {Peigne}}, \
  and\ \bibinfo {author} {\bibfnamefont {D.}~\bibnamefont {Schiff}},\ }\href
  {\doibase 10.1016/S0550-3213(96)00553-6} {\bibfield  {journal} {\bibinfo
  {journal} {Nucl. Phys. B}\ }\textbf {\bibinfo {volume} {483}},\ \bibinfo
  {pages} {291} (\bibinfo {year} {1997}{\natexlab{a}})},\ \Eprint
  {http://arxiv.org/abs/hep-ph/9607355} {arXiv:hep-ph/9607355} \BibitemShut
  {NoStop}%
\bibitem [{\citenamefont {Baier}\ \emph
  {et~al.}(1997{\natexlab{b}})\citenamefont {Baier}, \citenamefont
  {Dokshitzer}, \citenamefont {Mueller}, \citenamefont {Peigne},\ and\
  \citenamefont {Schiff}}]{Baier:1996sk}%
  \BibitemOpen
  \bibfield  {author} {\bibinfo {author} {\bibfnamefont {R.}~\bibnamefont
  {Baier}}, \bibinfo {author} {\bibfnamefont {Y.~L.}\ \bibnamefont
  {Dokshitzer}}, \bibinfo {author} {\bibfnamefont {A.~H.}\ \bibnamefont
  {Mueller}}, \bibinfo {author} {\bibfnamefont {S.}~\bibnamefont {Peigne}}, \
  and\ \bibinfo {author} {\bibfnamefont {D.}~\bibnamefont {Schiff}},\ }\href
  {\doibase 10.1016/S0550-3213(96)00581-0} {\bibfield  {journal} {\bibinfo
  {journal} {Nucl. Phys. B}\ }\textbf {\bibinfo {volume} {484}},\ \bibinfo
  {pages} {265} (\bibinfo {year} {1997}{\natexlab{b}})},\ \Eprint
  {http://arxiv.org/abs/hep-ph/9608322} {arXiv:hep-ph/9608322} \BibitemShut
  {NoStop}%
\bibitem [{\citenamefont {Zakharov}(1996)}]{Zakharov:1996fv}%
  \BibitemOpen
  \bibfield  {author} {\bibinfo {author} {\bibfnamefont {B.~G.}\ \bibnamefont
  {Zakharov}},\ }\href {\doibase 10.1134/1.567126} {\bibfield  {journal}
  {\bibinfo  {journal} {JETP Lett.}\ }\textbf {\bibinfo {volume} {63}},\
  \bibinfo {pages} {952} (\bibinfo {year} {1996})},\ \Eprint
  {http://arxiv.org/abs/hep-ph/9607440} {arXiv:hep-ph/9607440} \BibitemShut
  {NoStop}%
\bibitem [{\citenamefont {Zakharov}(1997)}]{Zakharov:1997uu}%
  \BibitemOpen
  \bibfield  {author} {\bibinfo {author} {\bibfnamefont {B.~G.}\ \bibnamefont
  {Zakharov}},\ }\href {\doibase 10.1134/1.567389} {\bibfield  {journal}
  {\bibinfo  {journal} {JETP Lett.}\ }\textbf {\bibinfo {volume} {65}},\
  \bibinfo {pages} {615} (\bibinfo {year} {1997})},\ \Eprint
  {http://arxiv.org/abs/hep-ph/9704255} {arXiv:hep-ph/9704255} \BibitemShut
  {NoStop}%
\bibitem [{\citenamefont {Gyulassy}\ \emph {et~al.}(2001)\citenamefont
  {Gyulassy}, \citenamefont {Levai},\ and\ \citenamefont
  {Vitev}}]{Gyulassy:2000er}%
  \BibitemOpen
  \bibfield  {author} {\bibinfo {author} {\bibfnamefont {M.}~\bibnamefont
  {Gyulassy}}, \bibinfo {author} {\bibfnamefont {P.}~\bibnamefont {Levai}}, \
  and\ \bibinfo {author} {\bibfnamefont {I.}~\bibnamefont {Vitev}},\ }\href
  {\doibase 10.1016/S0550-3213(00)00652-0} {\bibfield  {journal} {\bibinfo
  {journal} {Nucl. Phys. B}\ }\textbf {\bibinfo {volume} {594}},\ \bibinfo
  {pages} {371} (\bibinfo {year} {2001})},\ \Eprint
  {http://arxiv.org/abs/nucl-th/0006010} {arXiv:nucl-th/0006010} \BibitemShut
  {NoStop}%
\bibitem [{\citenamefont {Wiedemann}(2000)}]{Wiedemann:2000za}%
  \BibitemOpen
  \bibfield  {author} {\bibinfo {author} {\bibfnamefont {U.~A.}\ \bibnamefont
  {Wiedemann}},\ }\href {\doibase 10.1016/S0550-3213(00)00457-0} {\bibfield
  {journal} {\bibinfo  {journal} {Nucl. Phys. B}\ }\textbf {\bibinfo {volume}
  {588}},\ \bibinfo {pages} {303} (\bibinfo {year} {2000})},\ \Eprint
  {http://arxiv.org/abs/hep-ph/0005129} {arXiv:hep-ph/0005129} \BibitemShut
  {NoStop}%
\bibitem [{\citenamefont {Guo}\ and\ \citenamefont {Wang}(2000)}]{Guo:2000nz}%
  \BibitemOpen
  \bibfield  {author} {\bibinfo {author} {\bibfnamefont {X.-f.}\ \bibnamefont
  {Guo}}\ and\ \bibinfo {author} {\bibfnamefont {X.-N.}\ \bibnamefont {Wang}},\
  }\href {\doibase 10.1103/PhysRevLett.85.3591} {\bibfield  {journal} {\bibinfo
   {journal} {Phys. Rev. Lett.}\ }\textbf {\bibinfo {volume} {85}},\ \bibinfo
  {pages} {3591} (\bibinfo {year} {2000})},\ \Eprint
  {http://arxiv.org/abs/hep-ph/0005044} {arXiv:hep-ph/0005044} \BibitemShut
  {NoStop}%
\bibitem [{\citenamefont {Wang}\ and\ \citenamefont
  {Guo}(2001)}]{Wang:2001ifa}%
  \BibitemOpen
  \bibfield  {author} {\bibinfo {author} {\bibfnamefont {X.-N.}\ \bibnamefont
  {Wang}}\ and\ \bibinfo {author} {\bibfnamefont {X.-f.}\ \bibnamefont {Guo}},\
  }\href {\doibase 10.1016/S0375-9474(01)01130-7} {\bibfield  {journal}
  {\bibinfo  {journal} {Nucl. Phys. A}\ }\textbf {\bibinfo {volume} {696}},\
  \bibinfo {pages} {788} (\bibinfo {year} {2001})},\ \Eprint
  {http://arxiv.org/abs/hep-ph/0102230} {arXiv:hep-ph/0102230} \BibitemShut
  {NoStop}%
\bibitem [{\citenamefont {Arnold}\ \emph {et~al.}(2002)\citenamefont {Arnold},
  \citenamefont {Moore},\ and\ \citenamefont {Yaffe}}]{Arnold:2002ja}%
  \BibitemOpen
  \bibfield  {author} {\bibinfo {author} {\bibfnamefont {P.~B.}\ \bibnamefont
  {Arnold}}, \bibinfo {author} {\bibfnamefont {G.~D.}\ \bibnamefont {Moore}}, \
  and\ \bibinfo {author} {\bibfnamefont {L.~G.}\ \bibnamefont {Yaffe}},\ }\href
  {\doibase 10.1088/1126-6708/2002/06/030} {\bibfield  {journal} {\bibinfo
  {journal} {JHEP}\ }\textbf {\bibinfo {volume} {06}},\ \bibinfo {pages} {030}
  (\bibinfo {year} {2002})},\ \Eprint {http://arxiv.org/abs/hep-ph/0204343}
  {arXiv:hep-ph/0204343} \BibitemShut {NoStop}%
\bibitem [{\citenamefont {Arnold}\ \emph {et~al.}(2003)\citenamefont {Arnold},
  \citenamefont {Moore},\ and\ \citenamefont {Yaffe}}]{Arnold:2002zm}%
  \BibitemOpen
  \bibfield  {author} {\bibinfo {author} {\bibfnamefont {P.~B.}\ \bibnamefont
  {Arnold}}, \bibinfo {author} {\bibfnamefont {G.~D.}\ \bibnamefont {Moore}}, \
  and\ \bibinfo {author} {\bibfnamefont {L.~G.}\ \bibnamefont {Yaffe}},\ }\href
  {\doibase 10.1088/1126-6708/2003/01/030} {\bibfield  {journal} {\bibinfo
  {journal} {JHEP}\ }\textbf {\bibinfo {volume} {01}},\ \bibinfo {pages} {030}
  (\bibinfo {year} {2003})},\ \Eprint {http://arxiv.org/abs/hep-ph/0209353}
  {arXiv:hep-ph/0209353} \BibitemShut {NoStop}%
\bibitem [{\citenamefont {Salgado}\ and\ \citenamefont
  {Wiedemann}(2003)}]{Salgado:2003gb}%
  \BibitemOpen
  \bibfield  {author} {\bibinfo {author} {\bibfnamefont {C.~A.}\ \bibnamefont
  {Salgado}}\ and\ \bibinfo {author} {\bibfnamefont {U.~A.}\ \bibnamefont
  {Wiedemann}},\ }\href {\doibase 10.1103/PhysRevD.68.014008} {\bibfield
  {journal} {\bibinfo  {journal} {Phys. Rev. D}\ }\textbf {\bibinfo {volume}
  {68}},\ \bibinfo {pages} {014008} (\bibinfo {year} {2003})},\ \Eprint
  {http://arxiv.org/abs/hep-ph/0302184} {arXiv:hep-ph/0302184} \BibitemShut
  {NoStop}%
\bibitem [{\citenamefont {Liu}\ \emph {et~al.}(2006)\citenamefont {Liu},
  \citenamefont {Rajagopal},\ and\ \citenamefont {Wiedemann}}]{Liu:2006ug}%
  \BibitemOpen
  \bibfield  {author} {\bibinfo {author} {\bibfnamefont {H.}~\bibnamefont
  {Liu}}, \bibinfo {author} {\bibfnamefont {K.}~\bibnamefont {Rajagopal}}, \
  and\ \bibinfo {author} {\bibfnamefont {U.~A.}\ \bibnamefont {Wiedemann}},\
  }\href {\doibase 10.1103/PhysRevLett.97.182301} {\bibfield  {journal}
  {\bibinfo  {journal} {Phys. Rev. Lett.}\ }\textbf {\bibinfo {volume} {97}},\
  \bibinfo {pages} {182301} (\bibinfo {year} {2006})},\ \Eprint
  {http://arxiv.org/abs/hep-ph/0605178} {arXiv:hep-ph/0605178} \BibitemShut
  {NoStop}%
\bibitem [{\citenamefont {Qin}\ \emph {et~al.}(2008)\citenamefont {Qin},
  \citenamefont {Ruppert}, \citenamefont {Gale}, \citenamefont {Jeon},
  \citenamefont {Moore},\ and\ \citenamefont {Mustafa}}]{Qin:2007rn}%
  \BibitemOpen
  \bibfield  {author} {\bibinfo {author} {\bibfnamefont {G.-Y.}\ \bibnamefont
  {Qin}}, \bibinfo {author} {\bibfnamefont {J.}~\bibnamefont {Ruppert}},
  \bibinfo {author} {\bibfnamefont {C.}~\bibnamefont {Gale}}, \bibinfo {author}
  {\bibfnamefont {S.}~\bibnamefont {Jeon}}, \bibinfo {author} {\bibfnamefont
  {G.~D.}\ \bibnamefont {Moore}}, \ and\ \bibinfo {author} {\bibfnamefont
  {M.~G.}\ \bibnamefont {Mustafa}},\ }\href {\doibase
  10.1103/PhysRevLett.100.072301} {\bibfield  {journal} {\bibinfo  {journal}
  {Phys. Rev. Lett.}\ }\textbf {\bibinfo {volume} {100}},\ \bibinfo {pages}
  {072301} (\bibinfo {year} {2008})},\ \Eprint {http://arxiv.org/abs/0710.0605}
  {arXiv:0710.0605 [hep-ph]} \BibitemShut {NoStop}%
\bibitem [{\citenamefont {Armesto}\ \emph {et~al.}(2012)\citenamefont {Armesto}
  \emph {et~al.}}]{Armesto:2011ht}%
  \BibitemOpen
  \bibfield  {author} {\bibinfo {author} {\bibfnamefont {N.}~\bibnamefont
  {Armesto}} \emph {et~al.},\ }\href {\doibase 10.1103/PhysRevC.86.064904}
  {\bibfield  {journal} {\bibinfo  {journal} {Phys. Rev. C}\ }\textbf {\bibinfo
  {volume} {86}},\ \bibinfo {pages} {064904} (\bibinfo {year} {2012})},\
  \Eprint {http://arxiv.org/abs/1106.1106} {arXiv:1106.1106 [hep-ph]}
  \BibitemShut {NoStop}%
\bibitem [{\citenamefont {Mehtar-Tani}\ \emph {et~al.}(2011)\citenamefont
  {Mehtar-Tani}, \citenamefont {Salgado},\ and\ \citenamefont
  {Tywoniuk}}]{Mehtar-Tani:2010ebp}%
  \BibitemOpen
  \bibfield  {author} {\bibinfo {author} {\bibfnamefont {Y.}~\bibnamefont
  {Mehtar-Tani}}, \bibinfo {author} {\bibfnamefont {C.~A.}\ \bibnamefont
  {Salgado}}, \ and\ \bibinfo {author} {\bibfnamefont {K.}~\bibnamefont
  {Tywoniuk}},\ }\href {\doibase 10.1103/PhysRevLett.106.122002} {\bibfield
  {journal} {\bibinfo  {journal} {Phys. Rev. Lett.}\ }\textbf {\bibinfo
  {volume} {106}},\ \bibinfo {pages} {122002} (\bibinfo {year} {2011})},\
  \Eprint {http://arxiv.org/abs/1009.2965} {arXiv:1009.2965 [hep-ph]}
  \BibitemShut {NoStop}%
\bibitem [{\citenamefont {Mehtar-Tani}\ \emph
  {et~al.}(2012{\natexlab{a}})\citenamefont {Mehtar-Tani}, \citenamefont
  {Salgado},\ and\ \citenamefont {Tywoniuk}}]{Mehtar-Tani:2012mfa}%
  \BibitemOpen
  \bibfield  {author} {\bibinfo {author} {\bibfnamefont {Y.}~\bibnamefont
  {Mehtar-Tani}}, \bibinfo {author} {\bibfnamefont {C.~A.}\ \bibnamefont
  {Salgado}}, \ and\ \bibinfo {author} {\bibfnamefont {K.}~\bibnamefont
  {Tywoniuk}},\ }\href {\doibase 10.1007/JHEP10(2012)197} {\bibfield  {journal}
  {\bibinfo  {journal} {JHEP}\ }\textbf {\bibinfo {volume} {10}},\ \bibinfo
  {pages} {197} (\bibinfo {year} {2012}{\natexlab{a}})},\ \Eprint
  {http://arxiv.org/abs/1205.5739} {arXiv:1205.5739 [hep-ph]} \BibitemShut
  {NoStop}%
\bibitem [{\citenamefont {Casalderrey-Solana}\ and\ \citenamefont
  {Iancu}(2011)}]{Casalderrey-Solana:2011ule}%
  \BibitemOpen
  \bibfield  {author} {\bibinfo {author} {\bibfnamefont {J.}~\bibnamefont
  {Casalderrey-Solana}}\ and\ \bibinfo {author} {\bibfnamefont
  {E.}~\bibnamefont {Iancu}},\ }\href {\doibase 10.1007/JHEP08(2011)015}
  {\bibfield  {journal} {\bibinfo  {journal} {JHEP}\ }\textbf {\bibinfo
  {volume} {08}},\ \bibinfo {pages} {015} (\bibinfo {year} {2011})},\ \Eprint
  {http://arxiv.org/abs/1105.1760} {arXiv:1105.1760 [hep-ph]} \BibitemShut
  {NoStop}%
\bibitem [{\citenamefont {Casalderrey-Solana}\ \emph
  {et~al.}(2013)\citenamefont {Casalderrey-Solana}, \citenamefont
  {Mehtar-Tani}, \citenamefont {Salgado},\ and\ \citenamefont
  {Tywoniuk}}]{Casalderrey-Solana:2012evi}%
  \BibitemOpen
  \bibfield  {author} {\bibinfo {author} {\bibfnamefont {J.}~\bibnamefont
  {Casalderrey-Solana}}, \bibinfo {author} {\bibfnamefont {Y.}~\bibnamefont
  {Mehtar-Tani}}, \bibinfo {author} {\bibfnamefont {C.~A.}\ \bibnamefont
  {Salgado}}, \ and\ \bibinfo {author} {\bibfnamefont {K.}~\bibnamefont
  {Tywoniuk}},\ }\href {\doibase 10.1016/j.physletb.2013.07.046} {\bibfield
  {journal} {\bibinfo  {journal} {Phys. Lett. B}\ }\textbf {\bibinfo {volume}
  {725}},\ \bibinfo {pages} {357} (\bibinfo {year} {2013})},\ \Eprint
  {http://arxiv.org/abs/1210.7765} {arXiv:1210.7765 [hep-ph]} \BibitemShut
  {NoStop}%
\bibitem [{\citenamefont {Singh}\ and\ \citenamefont
  {Vaidya}(2024)}]{Singh:2024vwb}%
  \BibitemOpen
  \bibfield  {author} {\bibinfo {author} {\bibfnamefont {B.}~\bibnamefont
  {Singh}}\ and\ \bibinfo {author} {\bibfnamefont {V.}~\bibnamefont {Vaidya}},\
  }\href@noop {} {\  (\bibinfo {year} {2024})},\ \Eprint
  {http://arxiv.org/abs/2408.02753} {arXiv:2408.02753 [hep-ph]} \BibitemShut
  {NoStop}%
\bibitem [{\citenamefont {Idilbi}\ and\ \citenamefont
  {Majumder}(2009)}]{Idilbi:2008vm}%
  \BibitemOpen
  \bibfield  {author} {\bibinfo {author} {\bibfnamefont {A.}~\bibnamefont
  {Idilbi}}\ and\ \bibinfo {author} {\bibfnamefont {A.}~\bibnamefont
  {Majumder}},\ }\href {\doibase 10.1103/PhysRevD.80.054022} {\bibfield
  {journal} {\bibinfo  {journal} {Phys. Rev. D}\ }\textbf {\bibinfo {volume}
  {80}},\ \bibinfo {pages} {054022} (\bibinfo {year} {2009})},\ \Eprint
  {http://arxiv.org/abs/0808.1087} {arXiv:0808.1087 [hep-ph]} \BibitemShut
  {NoStop}%
\bibitem [{\citenamefont {D'Eramo}\ \emph {et~al.}(2011)\citenamefont
  {D'Eramo}, \citenamefont {Liu},\ and\ \citenamefont
  {Rajagopal}}]{DEramo:2010wup}%
  \BibitemOpen
  \bibfield  {author} {\bibinfo {author} {\bibfnamefont {F.}~\bibnamefont
  {D'Eramo}}, \bibinfo {author} {\bibfnamefont {H.}~\bibnamefont {Liu}}, \ and\
  \bibinfo {author} {\bibfnamefont {K.}~\bibnamefont {Rajagopal}},\ }\href
  {\doibase 10.1103/PhysRevD.84.065015} {\bibfield  {journal} {\bibinfo
  {journal} {Phys. Rev. D}\ }\textbf {\bibinfo {volume} {84}},\ \bibinfo
  {pages} {065015} (\bibinfo {year} {2011})},\ \Eprint
  {http://arxiv.org/abs/1006.1367} {arXiv:1006.1367 [hep-ph]} \BibitemShut
  {NoStop}%
\bibitem [{\citenamefont {Ovanesyan}\ and\ \citenamefont
  {Vitev}(2011)}]{Ovanesyan:2011xy}%
  \BibitemOpen
  \bibfield  {author} {\bibinfo {author} {\bibfnamefont {G.}~\bibnamefont
  {Ovanesyan}}\ and\ \bibinfo {author} {\bibfnamefont {I.}~\bibnamefont
  {Vitev}},\ }\href {\doibase 10.1007/JHEP06(2011)080} {\bibfield  {journal}
  {\bibinfo  {journal} {JHEP}\ }\textbf {\bibinfo {volume} {06}},\ \bibinfo
  {pages} {080} (\bibinfo {year} {2011})},\ \Eprint
  {http://arxiv.org/abs/1103.1074} {arXiv:1103.1074 [hep-ph]} \BibitemShut
  {NoStop}%
\bibitem [{\citenamefont {Dasgupta}\ \emph {et~al.}(2015)\citenamefont
  {Dasgupta}, \citenamefont {Dreyer}, \citenamefont {Salam},\ and\
  \citenamefont {Soyez}}]{Dasgupta:2014yra}%
  \BibitemOpen
  \bibfield  {author} {\bibinfo {author} {\bibfnamefont {M.}~\bibnamefont
  {Dasgupta}}, \bibinfo {author} {\bibfnamefont {F.}~\bibnamefont {Dreyer}},
  \bibinfo {author} {\bibfnamefont {G.~P.}\ \bibnamefont {Salam}}, \ and\
  \bibinfo {author} {\bibfnamefont {G.}~\bibnamefont {Soyez}},\ }\href
  {\doibase 10.1007/JHEP04(2015)039} {\bibfield  {journal} {\bibinfo  {journal}
  {JHEP}\ }\textbf {\bibinfo {volume} {04}},\ \bibinfo {pages} {039} (\bibinfo
  {year} {2015})},\ \Eprint {http://arxiv.org/abs/1411.5182} {arXiv:1411.5182
  [hep-ph]} \BibitemShut {NoStop}%
\bibitem [{\citenamefont {Kaufmann}\ \emph {et~al.}(2015)\citenamefont
  {Kaufmann}, \citenamefont {Mukherjee},\ and\ \citenamefont
  {Vogelsang}}]{Kaufmann:2015hma}%
  \BibitemOpen
  \bibfield  {author} {\bibinfo {author} {\bibfnamefont {T.}~\bibnamefont
  {Kaufmann}}, \bibinfo {author} {\bibfnamefont {A.}~\bibnamefont {Mukherjee}},
  \ and\ \bibinfo {author} {\bibfnamefont {W.}~\bibnamefont {Vogelsang}},\
  }\href {\doibase 10.1103/PhysRevD.92.054015} {\bibfield  {journal} {\bibinfo
  {journal} {Phys. Rev. D}\ }\textbf {\bibinfo {volume} {92}},\ \bibinfo
  {pages} {054015} (\bibinfo {year} {2015})},\ \bibinfo {note} {[Erratum:
  Phys.Rev.D 101, 079901 (2020)]},\ \Eprint {http://arxiv.org/abs/1506.01415}
  {arXiv:1506.01415 [hep-ph]} \BibitemShut {NoStop}%
\bibitem [{\citenamefont {Kang}\ \emph {et~al.}(2016)\citenamefont {Kang},
  \citenamefont {Ringer},\ and\ \citenamefont {Vitev}}]{Kang:2016mcy}%
  \BibitemOpen
  \bibfield  {author} {\bibinfo {author} {\bibfnamefont {Z.-B.}\ \bibnamefont
  {Kang}}, \bibinfo {author} {\bibfnamefont {F.}~\bibnamefont {Ringer}}, \ and\
  \bibinfo {author} {\bibfnamefont {I.}~\bibnamefont {Vitev}},\ }\href
  {\doibase 10.1007/JHEP10(2016)125} {\bibfield  {journal} {\bibinfo  {journal}
  {JHEP}\ }\textbf {\bibinfo {volume} {10}},\ \bibinfo {pages} {125} (\bibinfo
  {year} {2016})},\ \Eprint {http://arxiv.org/abs/1606.06732} {arXiv:1606.06732
  [hep-ph]} \BibitemShut {NoStop}%
\bibitem [{\citenamefont {Dai}\ \emph {et~al.}(2016)\citenamefont {Dai},
  \citenamefont {Kim},\ and\ \citenamefont {Leibovich}}]{Dai:2016hzf}%
  \BibitemOpen
  \bibfield  {author} {\bibinfo {author} {\bibfnamefont {L.}~\bibnamefont
  {Dai}}, \bibinfo {author} {\bibfnamefont {C.}~\bibnamefont {Kim}}, \ and\
  \bibinfo {author} {\bibfnamefont {A.~K.}\ \bibnamefont {Leibovich}},\ }\href
  {\doibase 10.1103/PhysRevD.94.114023} {\bibfield  {journal} {\bibinfo
  {journal} {Phys. Rev. D}\ }\textbf {\bibinfo {volume} {94}},\ \bibinfo
  {pages} {114023} (\bibinfo {year} {2016})},\ \Eprint
  {http://arxiv.org/abs/1606.07411} {arXiv:1606.07411 [hep-ph]} \BibitemShut
  {NoStop}%
\bibitem [{\citenamefont {van Beekveld}\ \emph {et~al.}(2024)\citenamefont {van
  Beekveld}, \citenamefont {Dasgupta}, \citenamefont {El-Menoufi},
  \citenamefont {Helliwell}, \citenamefont {Karlberg},\ and\ \citenamefont
  {Monni}}]{vanBeekveld:2024jnx}%
  \BibitemOpen
  \bibfield  {author} {\bibinfo {author} {\bibfnamefont {M.}~\bibnamefont {van
  Beekveld}}, \bibinfo {author} {\bibfnamefont {M.}~\bibnamefont {Dasgupta}},
  \bibinfo {author} {\bibfnamefont {B.~K.}\ \bibnamefont {El-Menoufi}},
  \bibinfo {author} {\bibfnamefont {J.}~\bibnamefont {Helliwell}}, \bibinfo
  {author} {\bibfnamefont {A.}~\bibnamefont {Karlberg}}, \ and\ \bibinfo
  {author} {\bibfnamefont {P.~F.}\ \bibnamefont {Monni}},\ }\href@noop {} {\
  (\bibinfo {year} {2024})},\ \Eprint {http://arxiv.org/abs/2402.05170}
  {arXiv:2402.05170 [hep-ph]} \BibitemShut {NoStop}%
\bibitem [{\citenamefont {Casalderrey-Solana}\ \emph
  {et~al.}(2014)\citenamefont {Casalderrey-Solana}, \citenamefont {Gulhan},
  \citenamefont {Milhano}, \citenamefont {Pablos},\ and\ \citenamefont
  {Rajagopal}}]{Casalderrey-Solana:2014bpa}%
  \BibitemOpen
  \bibfield  {author} {\bibinfo {author} {\bibfnamefont {J.}~\bibnamefont
  {Casalderrey-Solana}}, \bibinfo {author} {\bibfnamefont {D.~C.}\ \bibnamefont
  {Gulhan}}, \bibinfo {author} {\bibfnamefont {J.~G.}\ \bibnamefont {Milhano}},
  \bibinfo {author} {\bibfnamefont {D.}~\bibnamefont {Pablos}}, \ and\ \bibinfo
  {author} {\bibfnamefont {K.}~\bibnamefont {Rajagopal}},\ }\href {\doibase
  10.1007/JHEP09(2015)175} {\bibfield  {journal} {\bibinfo  {journal} {JHEP}\
  }\textbf {\bibinfo {volume} {10}},\ \bibinfo {pages} {019} (\bibinfo {year}
  {2014})},\ \bibinfo {note} {[Erratum: JHEP 09, 175 (2015)]},\ \Eprint
  {http://arxiv.org/abs/1405.3864} {arXiv:1405.3864 [hep-ph]} \BibitemShut
  {NoStop}%
\bibitem [{\citenamefont {Vaidya}\ and\ \citenamefont
  {Yao}(2020)}]{Vaidya:2020cyi}%
  \BibitemOpen
  \bibfield  {author} {\bibinfo {author} {\bibfnamefont {V.}~\bibnamefont
  {Vaidya}}\ and\ \bibinfo {author} {\bibfnamefont {X.}~\bibnamefont {Yao}},\
  }\href {\doibase 10.1007/JHEP10(2020)024} {\bibfield  {journal} {\bibinfo
  {journal} {JHEP}\ }\textbf {\bibinfo {volume} {10}},\ \bibinfo {pages} {024}
  (\bibinfo {year} {2020})},\ \Eprint {http://arxiv.org/abs/2004.11403}
  {arXiv:2004.11403 [hep-ph]} \BibitemShut {NoStop}%
\bibitem [{\citenamefont {Vaidya}(2021{\natexlab{a}})}]{Vaidya:2020lih}%
  \BibitemOpen
  \bibfield  {author} {\bibinfo {author} {\bibfnamefont {V.}~\bibnamefont
  {Vaidya}},\ }\href {\doibase 10.1007/JHEP11(2021)064} {\bibfield  {journal}
  {\bibinfo  {journal} {JHEP}\ }\textbf {\bibinfo {volume} {11}},\ \bibinfo
  {pages} {064} (\bibinfo {year} {2021}{\natexlab{a}})},\ \Eprint
  {http://arxiv.org/abs/2010.00028} {arXiv:2010.00028 [hep-ph]} \BibitemShut
  {NoStop}%
\bibitem [{\citenamefont {Vaidya}(2021{\natexlab{b}})}]{Vaidya:2021vxu}%
  \BibitemOpen
  \bibfield  {author} {\bibinfo {author} {\bibfnamefont {V.}~\bibnamefont
  {Vaidya}},\ }\href@noop {} {\  (\bibinfo {year} {2021}{\natexlab{b}})},\
  \Eprint {http://arxiv.org/abs/2107.00029} {arXiv:2107.00029 [hep-ph]}
  \BibitemShut {NoStop}%
\bibitem [{\citenamefont {Vaidya}(2021{\natexlab{c}})}]{Vaidya:2021mly}%
  \BibitemOpen
  \bibfield  {author} {\bibinfo {author} {\bibfnamefont {V.}~\bibnamefont
  {Vaidya}},\ }\href@noop {} {\  (\bibinfo {year} {2021}{\natexlab{c}})},\
  \Eprint {http://arxiv.org/abs/2109.11568} {arXiv:2109.11568 [hep-ph]}
  \BibitemShut {NoStop}%
\bibitem [{\citenamefont {Andersson}\ \emph {et~al.}(1989)\citenamefont
  {Andersson}, \citenamefont {Gustafson}, \citenamefont {Lonnblad},\ and\
  \citenamefont {Pettersson}}]{Andersson:1988gp}%
  \BibitemOpen
  \bibfield  {author} {\bibinfo {author} {\bibfnamefont {B.}~\bibnamefont
  {Andersson}}, \bibinfo {author} {\bibfnamefont {G.}~\bibnamefont
  {Gustafson}}, \bibinfo {author} {\bibfnamefont {L.}~\bibnamefont {Lonnblad}},
  \ and\ \bibinfo {author} {\bibfnamefont {U.}~\bibnamefont {Pettersson}},\
  }\href {\doibase 10.1007/BF01550942} {\bibfield  {journal} {\bibinfo
  {journal} {Z. Phys. C}\ }\textbf {\bibinfo {volume} {43}},\ \bibinfo {pages}
  {625} (\bibinfo {year} {1989})}\BibitemShut {NoStop}%
\bibitem [{\citenamefont {Mehtar-Tani}\ and\ \citenamefont
  {Tywoniuk}(2018)}]{Mehtar-Tani:2017web}%
  \BibitemOpen
  \bibfield  {author} {\bibinfo {author} {\bibfnamefont {Y.}~\bibnamefont
  {Mehtar-Tani}}\ and\ \bibinfo {author} {\bibfnamefont {K.}~\bibnamefont
  {Tywoniuk}},\ }\href {\doibase 10.1103/PhysRevD.98.051501} {\bibfield
  {journal} {\bibinfo  {journal} {Phys. Rev. D}\ }\textbf {\bibinfo {volume}
  {98}},\ \bibinfo {pages} {051501} (\bibinfo {year} {2018})},\ \Eprint
  {http://arxiv.org/abs/1707.07361} {arXiv:1707.07361 [hep-ph]} \BibitemShut
  {NoStop}%
\bibitem [{\citenamefont {Caucal}\ \emph {et~al.}(2018)\citenamefont {Caucal},
  \citenamefont {Iancu}, \citenamefont {Mueller},\ and\ \citenamefont
  {Soyez}}]{Caucal:2018dla}%
  \BibitemOpen
  \bibfield  {author} {\bibinfo {author} {\bibfnamefont {P.}~\bibnamefont
  {Caucal}}, \bibinfo {author} {\bibfnamefont {E.}~\bibnamefont {Iancu}},
  \bibinfo {author} {\bibfnamefont {A.~H.}\ \bibnamefont {Mueller}}, \ and\
  \bibinfo {author} {\bibfnamefont {G.}~\bibnamefont {Soyez}},\ }\href
  {\doibase 10.1103/PhysRevLett.120.232001} {\bibfield  {journal} {\bibinfo
  {journal} {Phys. Rev. Lett.}\ }\textbf {\bibinfo {volume} {120}},\ \bibinfo
  {pages} {232001} (\bibinfo {year} {2018})},\ \Eprint
  {http://arxiv.org/abs/1801.09703} {arXiv:1801.09703 [hep-ph]} \BibitemShut
  {NoStop}%
\bibitem [{\citenamefont {Mehtar-Tani}\ \emph {et~al.}()\citenamefont
  {Mehtar-Tani}, \citenamefont {Ringer}, \citenamefont {Singh},\ and\
  \citenamefont {Vaidya}}]{LongDerive}%
  \BibitemOpen
  \bibfield  {author} {\bibinfo {author} {\bibfnamefont {Y.}~\bibnamefont
  {Mehtar-Tani}}, \bibinfo {author} {\bibfnamefont {F.}~\bibnamefont {Ringer}},
  \bibinfo {author} {\bibfnamefont {B.}~\bibnamefont {Singh}}, \ and\ \bibinfo
  {author} {\bibfnamefont {V.}~\bibnamefont {Vaidya}},\ }\href@noop {} {\
  }\Eprint {http://arxiv.org/abs/to appear} {to appear} \BibitemShut {NoStop}%
\bibitem [{\citenamefont {Kang}\ \emph {et~al.}(2017)\citenamefont {Kang},
  \citenamefont {Ringer},\ and\ \citenamefont {Vitev}}]{Kang:2017frl}%
  \BibitemOpen
  \bibfield  {author} {\bibinfo {author} {\bibfnamefont {Z.-B.}\ \bibnamefont
  {Kang}}, \bibinfo {author} {\bibfnamefont {F.}~\bibnamefont {Ringer}}, \ and\
  \bibinfo {author} {\bibfnamefont {I.}~\bibnamefont {Vitev}},\ }\href
  {\doibase 10.1016/j.physletb.2017.03.067} {\bibfield  {journal} {\bibinfo
  {journal} {Phys. Lett. B}\ }\textbf {\bibinfo {volume} {769}},\ \bibinfo
  {pages} {242} (\bibinfo {year} {2017})},\ \Eprint
  {http://arxiv.org/abs/1701.05839} {arXiv:1701.05839 [hep-ph]} \BibitemShut
  {NoStop}%
\bibitem [{\citenamefont {He}\ \emph {et~al.}(2019)\citenamefont {He},
  \citenamefont {Pang},\ and\ \citenamefont {Wang}}]{He:2018gks}%
  \BibitemOpen
  \bibfield  {author} {\bibinfo {author} {\bibfnamefont {Y.}~\bibnamefont
  {He}}, \bibinfo {author} {\bibfnamefont {L.-G.}\ \bibnamefont {Pang}}, \ and\
  \bibinfo {author} {\bibfnamefont {X.-N.}\ \bibnamefont {Wang}},\ }\href
  {\doibase 10.1103/PhysRevLett.122.252302} {\bibfield  {journal} {\bibinfo
  {journal} {Phys. Rev. Lett.}\ }\textbf {\bibinfo {volume} {122}},\ \bibinfo
  {pages} {252302} (\bibinfo {year} {2019})},\ \Eprint
  {http://arxiv.org/abs/1808.05310} {arXiv:1808.05310 [hep-ph]} \BibitemShut
  {NoStop}%
\bibitem [{\citenamefont {Qiu}\ \emph {et~al.}(2019)\citenamefont {Qiu},
  \citenamefont {Ringer}, \citenamefont {Sato},\ and\ \citenamefont
  {Zurita}}]{Qiu:2019sfj}%
  \BibitemOpen
  \bibfield  {author} {\bibinfo {author} {\bibfnamefont {J.-W.}\ \bibnamefont
  {Qiu}}, \bibinfo {author} {\bibfnamefont {F.}~\bibnamefont {Ringer}},
  \bibinfo {author} {\bibfnamefont {N.}~\bibnamefont {Sato}}, \ and\ \bibinfo
  {author} {\bibfnamefont {P.}~\bibnamefont {Zurita}},\ }\href {\doibase
  10.1103/PhysRevLett.122.252301} {\bibfield  {journal} {\bibinfo  {journal}
  {Phys. Rev. Lett.}\ }\textbf {\bibinfo {volume} {122}},\ \bibinfo {pages}
  {252301} (\bibinfo {year} {2019})},\ \Eprint
  {http://arxiv.org/abs/1903.01993} {arXiv:1903.01993 [hep-ph]} \BibitemShut
  {NoStop}%
\bibitem [{\citenamefont {Becher}\ \emph {et~al.}(2016)\citenamefont {Becher},
  \citenamefont {Neubert}, \citenamefont {Rothen},\ and\ \citenamefont
  {Shao}}]{Becher:2015hka}%
  \BibitemOpen
  \bibfield  {author} {\bibinfo {author} {\bibfnamefont {T.}~\bibnamefont
  {Becher}}, \bibinfo {author} {\bibfnamefont {M.}~\bibnamefont {Neubert}},
  \bibinfo {author} {\bibfnamefont {L.}~\bibnamefont {Rothen}}, \ and\ \bibinfo
  {author} {\bibfnamefont {D.~Y.}\ \bibnamefont {Shao}},\ }\href {\doibase
  10.1103/PhysRevLett.116.192001} {\bibfield  {journal} {\bibinfo  {journal}
  {Phys. Rev. Lett.}\ }\textbf {\bibinfo {volume} {116}},\ \bibinfo {pages}
  {192001} (\bibinfo {year} {2016})},\ \Eprint
  {http://arxiv.org/abs/1508.06645} {arXiv:1508.06645 [hep-ph]} \BibitemShut
  {NoStop}%
\bibitem [{\citenamefont {Dasgupta}\ and\ \citenamefont
  {Salam}(2001)}]{Dasgupta:2001sh}%
  \BibitemOpen
  \bibfield  {author} {\bibinfo {author} {\bibfnamefont {M.}~\bibnamefont
  {Dasgupta}}\ and\ \bibinfo {author} {\bibfnamefont {G.~P.}\ \bibnamefont
  {Salam}},\ }\href {\doibase 10.1016/S0370-2693(01)00725-0} {\bibfield
  {journal} {\bibinfo  {journal} {Phys. Lett. B}\ }\textbf {\bibinfo {volume}
  {512}},\ \bibinfo {pages} {323} (\bibinfo {year} {2001})},\ \Eprint
  {http://arxiv.org/abs/hep-ph/0104277} {arXiv:hep-ph/0104277} \BibitemShut
  {NoStop}%
\bibitem [{\citenamefont {Larkoski}\ \emph {et~al.}(2015)\citenamefont
  {Larkoski}, \citenamefont {Moult},\ and\ \citenamefont
  {Neill}}]{Larkoski:2015zka}%
  \BibitemOpen
  \bibfield  {author} {\bibinfo {author} {\bibfnamefont {A.~J.}\ \bibnamefont
  {Larkoski}}, \bibinfo {author} {\bibfnamefont {I.}~\bibnamefont {Moult}}, \
  and\ \bibinfo {author} {\bibfnamefont {D.}~\bibnamefont {Neill}},\ }\href
  {\doibase 10.1007/JHEP09(2015)143} {\bibfield  {journal} {\bibinfo  {journal}
  {JHEP}\ }\textbf {\bibinfo {volume} {09}},\ \bibinfo {pages} {143} (\bibinfo
  {year} {2015})},\ \Eprint {http://arxiv.org/abs/1501.04596} {arXiv:1501.04596
  [hep-ph]} \BibitemShut {NoStop}%
\bibitem [{\citenamefont {Balitsky}(1996)}]{Balitsky:1995ub}%
  \BibitemOpen
  \bibfield  {author} {\bibinfo {author} {\bibfnamefont {I.}~\bibnamefont
  {Balitsky}},\ }\href {\doibase 10.1016/0550-3213(95)00638-9} {\bibfield
  {journal} {\bibinfo  {journal} {Nucl. Phys. B}\ }\textbf {\bibinfo {volume}
  {463}},\ \bibinfo {pages} {99} (\bibinfo {year} {1996})},\ \Eprint
  {http://arxiv.org/abs/hep-ph/9509348} {arXiv:hep-ph/9509348} \BibitemShut
  {NoStop}%
\bibitem [{\citenamefont {Chiu}\ \emph {et~al.}(2012)\citenamefont {Chiu},
  \citenamefont {Jain}, \citenamefont {Neill},\ and\ \citenamefont
  {Rothstein}}]{Chiu:2012ir}%
  \BibitemOpen
  \bibfield  {author} {\bibinfo {author} {\bibfnamefont {J.-Y.}\ \bibnamefont
  {Chiu}}, \bibinfo {author} {\bibfnamefont {A.}~\bibnamefont {Jain}}, \bibinfo
  {author} {\bibfnamefont {D.}~\bibnamefont {Neill}}, \ and\ \bibinfo {author}
  {\bibfnamefont {I.~Z.}\ \bibnamefont {Rothstein}},\ }\href {\doibase
  10.1007/JHEP05(2012)084} {\bibfield  {journal} {\bibinfo  {journal} {JHEP}\
  }\textbf {\bibinfo {volume} {05}},\ \bibinfo {pages} {084} (\bibinfo {year}
  {2012})},\ \Eprint {http://arxiv.org/abs/1202.0814} {arXiv:1202.0814
  [hep-ph]} \BibitemShut {NoStop}%
\bibitem [{\citenamefont {Iancu}(2014)}]{Iancu:2014kga}%
  \BibitemOpen
  \bibfield  {author} {\bibinfo {author} {\bibfnamefont {E.}~\bibnamefont
  {Iancu}},\ }\href {\doibase 10.1007/JHEP10(2014)095} {\bibfield  {journal}
  {\bibinfo  {journal} {JHEP}\ }\textbf {\bibinfo {volume} {10}},\ \bibinfo
  {pages} {095} (\bibinfo {year} {2014})},\ \Eprint
  {http://arxiv.org/abs/1403.1996} {arXiv:1403.1996 [hep-ph]} \BibitemShut
  {NoStop}%
\bibitem [{\citenamefont {Blaizot}\ \emph {et~al.}(2013)\citenamefont
  {Blaizot}, \citenamefont {Dominguez}, \citenamefont {Iancu},\ and\
  \citenamefont {Mehtar-Tani}}]{Blaizot:2012fh}%
  \BibitemOpen
  \bibfield  {author} {\bibinfo {author} {\bibfnamefont {J.-P.}\ \bibnamefont
  {Blaizot}}, \bibinfo {author} {\bibfnamefont {F.}~\bibnamefont {Dominguez}},
  \bibinfo {author} {\bibfnamefont {E.}~\bibnamefont {Iancu}}, \ and\ \bibinfo
  {author} {\bibfnamefont {Y.}~\bibnamefont {Mehtar-Tani}},\ }\href {\doibase
  10.1007/JHEP01(2013)143} {\bibfield  {journal} {\bibinfo  {journal} {JHEP}\
  }\textbf {\bibinfo {volume} {01}},\ \bibinfo {pages} {143} (\bibinfo {year}
  {2013})},\ \Eprint {http://arxiv.org/abs/1209.4585} {arXiv:1209.4585
  [hep-ph]} \BibitemShut {NoStop}%
\bibitem [{\citenamefont {Blaizot}\ and\ \citenamefont
  {Mehtar-Tani}(2014)}]{Blaizot:2014bha}%
  \BibitemOpen
  \bibfield  {author} {\bibinfo {author} {\bibfnamefont {J.-P.}\ \bibnamefont
  {Blaizot}}\ and\ \bibinfo {author} {\bibfnamefont {Y.}~\bibnamefont
  {Mehtar-Tani}},\ }\href {\doibase 10.1016/j.nuclphysa.2014.05.018} {\bibfield
   {journal} {\bibinfo  {journal} {Nucl. Phys. A}\ }\textbf {\bibinfo {volume}
  {929}},\ \bibinfo {pages} {202} (\bibinfo {year} {2014})},\ \Eprint
  {http://arxiv.org/abs/1403.2323} {arXiv:1403.2323 [hep-ph]} \BibitemShut
  {NoStop}%
\bibitem [{\citenamefont {Blaizot}\ \emph {et~al.}(2014)\citenamefont
  {Blaizot}, \citenamefont {Dominguez}, \citenamefont {Iancu},\ and\
  \citenamefont {Mehtar-Tani}}]{Blaizot:2013vha}%
  \BibitemOpen
  \bibfield  {author} {\bibinfo {author} {\bibfnamefont {J.-P.}\ \bibnamefont
  {Blaizot}}, \bibinfo {author} {\bibfnamefont {F.}~\bibnamefont {Dominguez}},
  \bibinfo {author} {\bibfnamefont {E.}~\bibnamefont {Iancu}}, \ and\ \bibinfo
  {author} {\bibfnamefont {Y.}~\bibnamefont {Mehtar-Tani}},\ }\href {\doibase
  10.1007/JHEP06(2014)075} {\bibfield  {journal} {\bibinfo  {journal} {JHEP}\
  }\textbf {\bibinfo {volume} {06}},\ \bibinfo {pages} {075} (\bibinfo {year}
  {2014})},\ \Eprint {http://arxiv.org/abs/1311.5823} {arXiv:1311.5823
  [hep-ph]} \BibitemShut {NoStop}%
\bibitem [{\citenamefont {Rothstein}\ and\ \citenamefont
  {Stewart}(2016)}]{Rothstein:2016bsq}%
  \BibitemOpen
  \bibfield  {author} {\bibinfo {author} {\bibfnamefont {I.~Z.}\ \bibnamefont
  {Rothstein}}\ and\ \bibinfo {author} {\bibfnamefont {I.~W.}\ \bibnamefont
  {Stewart}},\ }\href {\doibase 10.1007/JHEP08(2016)025} {\bibfield  {journal}
  {\bibinfo  {journal} {JHEP}\ }\textbf {\bibinfo {volume} {08}},\ \bibinfo
  {pages} {025} (\bibinfo {year} {2016})},\ \Eprint
  {http://arxiv.org/abs/1601.04695} {arXiv:1601.04695 [hep-ph]} \BibitemShut
  {NoStop}%
\bibitem [{\citenamefont {Stewart}\ and\ \citenamefont
  {Vaidya}(2024)}]{Stewart:2023lwz}%
  \BibitemOpen
  \bibfield  {author} {\bibinfo {author} {\bibfnamefont {I.~W.}\ \bibnamefont
  {Stewart}}\ and\ \bibinfo {author} {\bibfnamefont {V.}~\bibnamefont
  {Vaidya}},\ }\href {\doibase 10.1103/PhysRevD.110.L011504} {\bibfield
  {journal} {\bibinfo  {journal} {Phys. Rev. D}\ }\textbf {\bibinfo {volume}
  {110}},\ \bibinfo {pages} {L011504} (\bibinfo {year} {2024})},\ \Eprint
  {http://arxiv.org/abs/2305.16393} {arXiv:2305.16393 [hep-ph]} \BibitemShut
  {NoStop}%
\bibitem [{\citenamefont {Balitsky}\ and\ \citenamefont
  {Lipatov}(1978)}]{Balitsky:1978ic}%
  \BibitemOpen
  \bibfield  {author} {\bibinfo {author} {\bibfnamefont {I.~I.}\ \bibnamefont
  {Balitsky}}\ and\ \bibinfo {author} {\bibfnamefont {L.~N.}\ \bibnamefont
  {Lipatov}},\ }\href@noop {} {\bibfield  {journal} {\bibinfo  {journal} {Sov.
  J. Nucl. Phys.}\ }\textbf {\bibinfo {volume} {28}},\ \bibinfo {pages} {822}
  (\bibinfo {year} {1978})}\BibitemShut {NoStop}%
\bibitem [{\citenamefont {Kuraev}\ \emph {et~al.}(1977)\citenamefont {Kuraev},
  \citenamefont {Lipatov},\ and\ \citenamefont {Fadin}}]{Kuraev:1977fs}%
  \BibitemOpen
  \bibfield  {author} {\bibinfo {author} {\bibfnamefont {E.~A.}\ \bibnamefont
  {Kuraev}}, \bibinfo {author} {\bibfnamefont {L.~N.}\ \bibnamefont {Lipatov}},
  \ and\ \bibinfo {author} {\bibfnamefont {V.~S.}\ \bibnamefont {Fadin}},\
  }\href@noop {} {\bibfield  {journal} {\bibinfo  {journal} {Sov. Phys. JETP}\
  }\textbf {\bibinfo {volume} {45}},\ \bibinfo {pages} {199} (\bibinfo {year}
  {1977})}\BibitemShut {NoStop}%
\bibitem [{\citenamefont {Liou}\ \emph {et~al.}(2013)\citenamefont {Liou},
  \citenamefont {Mueller},\ and\ \citenamefont {Wu}}]{Liou:2013qya}%
  \BibitemOpen
  \bibfield  {author} {\bibinfo {author} {\bibfnamefont {T.}~\bibnamefont
  {Liou}}, \bibinfo {author} {\bibfnamefont {A.~H.}\ \bibnamefont {Mueller}}, \
  and\ \bibinfo {author} {\bibfnamefont {B.}~\bibnamefont {Wu}},\ }\href
  {\doibase 10.1016/j.nuclphysa.2013.08.005} {\bibfield  {journal} {\bibinfo
  {journal} {Nucl. Phys. A}\ }\textbf {\bibinfo {volume} {916}},\ \bibinfo
  {pages} {102} (\bibinfo {year} {2013})},\ \Eprint
  {http://arxiv.org/abs/1304.7677} {arXiv:1304.7677 [hep-ph]} \BibitemShut
  {NoStop}%
\bibitem [{\citenamefont {Mehtar-Tani}\ \emph {et~al.}(2013)\citenamefont
  {Mehtar-Tani}, \citenamefont {Milhano},\ and\ \citenamefont
  {Tywoniuk}}]{Mehtar-Tani:2013pia}%
  \BibitemOpen
  \bibfield  {author} {\bibinfo {author} {\bibfnamefont {Y.}~\bibnamefont
  {Mehtar-Tani}}, \bibinfo {author} {\bibfnamefont {J.~G.}\ \bibnamefont
  {Milhano}}, \ and\ \bibinfo {author} {\bibfnamefont {K.}~\bibnamefont
  {Tywoniuk}},\ }\href {\doibase 10.1142/S0217751X13400137} {\bibfield
  {journal} {\bibinfo  {journal} {Int. J. Mod. Phys. A}\ }\textbf {\bibinfo
  {volume} {28}},\ \bibinfo {pages} {1340013} (\bibinfo {year} {2013})},\
  \Eprint {http://arxiv.org/abs/1302.2579} {arXiv:1302.2579 [hep-ph]}
  \BibitemShut {NoStop}%
\bibitem [{\citenamefont {Arnold}\ \emph {et~al.}(2022)\citenamefont {Arnold},
  \citenamefont {Gorda},\ and\ \citenamefont {Iqbal}}]{Arnold:2021pin}%
  \BibitemOpen
  \bibfield  {author} {\bibinfo {author} {\bibfnamefont {P.}~\bibnamefont
  {Arnold}}, \bibinfo {author} {\bibfnamefont {T.}~\bibnamefont {Gorda}}, \
  and\ \bibinfo {author} {\bibfnamefont {S.}~\bibnamefont {Iqbal}},\ }\href
  {\doibase 10.1007/JHEP04(2022)085} {\bibfield  {journal} {\bibinfo  {journal}
  {JHEP}\ }\textbf {\bibinfo {volume} {04}},\ \bibinfo {pages} {085} (\bibinfo
  {year} {2022})},\ \Eprint {http://arxiv.org/abs/2112.05161} {arXiv:2112.05161
  [hep-ph]} \BibitemShut {NoStop}%
\bibitem [{\citenamefont {Gyulassy}\ \emph {et~al.}(2000)\citenamefont
  {Gyulassy}, \citenamefont {Levai},\ and\ \citenamefont
  {Vitev}}]{Gyulassy:2000fs}%
  \BibitemOpen
  \bibfield  {author} {\bibinfo {author} {\bibfnamefont {M.}~\bibnamefont
  {Gyulassy}}, \bibinfo {author} {\bibfnamefont {P.}~\bibnamefont {Levai}}, \
  and\ \bibinfo {author} {\bibfnamefont {I.}~\bibnamefont {Vitev}},\ }\href
  {\doibase 10.1103/PhysRevLett.85.5535} {\bibfield  {journal} {\bibinfo
  {journal} {Phys. Rev. Lett.}\ }\textbf {\bibinfo {volume} {85}},\ \bibinfo
  {pages} {5535} (\bibinfo {year} {2000})},\ \Eprint
  {http://arxiv.org/abs/nucl-th/0005032} {arXiv:nucl-th/0005032} \BibitemShut
  {NoStop}%
\bibitem [{\citenamefont {de~Florian}\ and\ \citenamefont
  {Vogelsang}(2007)}]{deFlorian:2007fv}%
  \BibitemOpen
  \bibfield  {author} {\bibinfo {author} {\bibfnamefont {D.}~\bibnamefont
  {de~Florian}}\ and\ \bibinfo {author} {\bibfnamefont {W.}~\bibnamefont
  {Vogelsang}},\ }\href {\doibase 10.1103/PhysRevD.76.074031} {\bibfield
  {journal} {\bibinfo  {journal} {Phys. Rev. D}\ }\textbf {\bibinfo {volume}
  {76}},\ \bibinfo {pages} {074031} (\bibinfo {year} {2007})},\ \Eprint
  {http://arxiv.org/abs/0704.1677} {arXiv:0704.1677 [hep-ph]} \BibitemShut
  {NoStop}%
\bibitem [{\citenamefont {Dai}\ \emph {et~al.}(2017)\citenamefont {Dai},
  \citenamefont {Kim},\ and\ \citenamefont {Leibovich}}]{Dai:2017dpc}%
  \BibitemOpen
  \bibfield  {author} {\bibinfo {author} {\bibfnamefont {L.}~\bibnamefont
  {Dai}}, \bibinfo {author} {\bibfnamefont {C.}~\bibnamefont {Kim}}, \ and\
  \bibinfo {author} {\bibfnamefont {A.~K.}\ \bibnamefont {Leibovich}},\ }\href
  {\doibase 10.1103/PhysRevD.95.074003} {\bibfield  {journal} {\bibinfo
  {journal} {Phys. Rev. D}\ }\textbf {\bibinfo {volume} {95}},\ \bibinfo
  {pages} {074003} (\bibinfo {year} {2017})},\ \Eprint
  {http://arxiv.org/abs/1701.05660} {arXiv:1701.05660 [hep-ph]} \BibitemShut
  {NoStop}%
\bibitem [{\citenamefont {Liu}\ \emph {et~al.}(2017)\citenamefont {Liu},
  \citenamefont {Moch},\ and\ \citenamefont {Ringer}}]{Liu:2017pbb}%
  \BibitemOpen
  \bibfield  {author} {\bibinfo {author} {\bibfnamefont {X.}~\bibnamefont
  {Liu}}, \bibinfo {author} {\bibfnamefont {S.-O.}\ \bibnamefont {Moch}}, \
  and\ \bibinfo {author} {\bibfnamefont {F.}~\bibnamefont {Ringer}},\ }\href
  {\doibase 10.1103/PhysRevLett.119.212001} {\bibfield  {journal} {\bibinfo
  {journal} {Phys. Rev. Lett.}\ }\textbf {\bibinfo {volume} {119}},\ \bibinfo
  {pages} {212001} (\bibinfo {year} {2017})},\ \Eprint
  {http://arxiv.org/abs/1708.04641} {arXiv:1708.04641 [hep-ph]} \BibitemShut
  {NoStop}%
\bibitem [{\citenamefont {Neill}\ \emph {et~al.}(2021)\citenamefont {Neill},
  \citenamefont {Ringer},\ and\ \citenamefont {Sato}}]{Neill:2021std}%
  \BibitemOpen
  \bibfield  {author} {\bibinfo {author} {\bibfnamefont {D.}~\bibnamefont
  {Neill}}, \bibinfo {author} {\bibfnamefont {F.}~\bibnamefont {Ringer}}, \
  and\ \bibinfo {author} {\bibfnamefont {N.}~\bibnamefont {Sato}},\ }\href
  {\doibase 10.1007/JHEP07(2021)041} {\bibfield  {journal} {\bibinfo  {journal}
  {JHEP}\ }\textbf {\bibinfo {volume} {07}},\ \bibinfo {pages} {041} (\bibinfo
  {year} {2021})},\ \Eprint {http://arxiv.org/abs/2103.16573} {arXiv:2103.16573
  [hep-ph]} \BibitemShut {NoStop}%
\bibitem [{\citenamefont {Mehtar-Tani}(2007)}]{Mehtar-Tani:2006vpj}%
  \BibitemOpen
  \bibfield  {author} {\bibinfo {author} {\bibfnamefont {Y.}~\bibnamefont
  {Mehtar-Tani}},\ }\href {\doibase 10.1103/PhysRevC.75.034908} {\bibfield
  {journal} {\bibinfo  {journal} {Phys. Rev. C}\ }\textbf {\bibinfo {volume}
  {75}},\ \bibinfo {pages} {034908} (\bibinfo {year} {2007})},\ \Eprint
  {http://arxiv.org/abs/hep-ph/0606236} {arXiv:hep-ph/0606236} \BibitemShut
  {NoStop}%
\bibitem [{\citenamefont {Blaizot}\ and\ \citenamefont
  {Mehtar-Tani}(2015)}]{Blaizot:2015lma}%
  \BibitemOpen
  \bibfield  {author} {\bibinfo {author} {\bibfnamefont {J.-P.}\ \bibnamefont
  {Blaizot}}\ and\ \bibinfo {author} {\bibfnamefont {Y.}~\bibnamefont
  {Mehtar-Tani}},\ }\href {\doibase 10.1142/S021830131530012X} {\bibfield
  {journal} {\bibinfo  {journal} {Int. J. Mod. Phys. E}\ }\textbf {\bibinfo
  {volume} {24}},\ \bibinfo {pages} {1530012} (\bibinfo {year} {2015})},\
  \Eprint {http://arxiv.org/abs/1503.05958} {arXiv:1503.05958 [hep-ph]}
  \BibitemShut {NoStop}%
\bibitem [{\citenamefont {Gelis}\ \emph {et~al.}(2010)\citenamefont {Gelis},
  \citenamefont {Iancu}, \citenamefont {Jalilian-Marian},\ and\ \citenamefont
  {Venugopalan}}]{Gelis:2010nm}%
  \BibitemOpen
  \bibfield  {author} {\bibinfo {author} {\bibfnamefont {F.}~\bibnamefont
  {Gelis}}, \bibinfo {author} {\bibfnamefont {E.}~\bibnamefont {Iancu}},
  \bibinfo {author} {\bibfnamefont {J.}~\bibnamefont {Jalilian-Marian}}, \ and\
  \bibinfo {author} {\bibfnamefont {R.}~\bibnamefont {Venugopalan}},\ }\href
  {\doibase 10.1146/annurev.nucl.010909.083629} {\bibfield  {journal} {\bibinfo
   {journal} {Ann. Rev. Nucl. Part. Sci.}\ }\textbf {\bibinfo {volume} {60}},\
  \bibinfo {pages} {463} (\bibinfo {year} {2010})},\ \Eprint
  {http://arxiv.org/abs/1002.0333} {arXiv:1002.0333 [hep-ph]} \BibitemShut
  {NoStop}%
\bibitem [{\citenamefont {Mehtar-Tani}\ \emph
  {et~al.}(2012{\natexlab{b}})\citenamefont {Mehtar-Tani}, \citenamefont
  {Salgado},\ and\ \citenamefont {Tywoniuk}}]{Mehtar-Tani:2011lic}%
  \BibitemOpen
  \bibfield  {author} {\bibinfo {author} {\bibfnamefont {Y.}~\bibnamefont
  {Mehtar-Tani}}, \bibinfo {author} {\bibfnamefont {C.~A.}\ \bibnamefont
  {Salgado}}, \ and\ \bibinfo {author} {\bibfnamefont {K.}~\bibnamefont
  {Tywoniuk}},\ }\href {\doibase 10.1007/JHEP04(2012)064} {\bibfield  {journal}
  {\bibinfo  {journal} {JHEP}\ }\textbf {\bibinfo {volume} {04}},\ \bibinfo
  {pages} {064} (\bibinfo {year} {2012}{\natexlab{b}})},\ \Eprint
  {http://arxiv.org/abs/1112.5031} {arXiv:1112.5031 [hep-ph]} \BibitemShut
  {NoStop}%
\end{thebibliography}%

\end{document}